\documentclass[aps,prb,superscriptaddress,twocolumn]{revtex4-2}

\usepackage{amsmath,amssymb}
\usepackage{graphicx,stackengine,scalerel}
\usepackage[colorlinks=true,urlcolor=blue]{hyperref}
\usepackage{dcolumn}
\usepackage{multirow,booktabs,hhline}
\usepackage{bm}
\usepackage[normalem]{ulem}
\usepackage{stmaryrd}
\usepackage[dvipsnames]{xcolor}
\definecolor{psfill}{RGB}{218,73,253}
\definecolor{nsfill}{RGB}{246,173,112}
\definecolor{anisgreen}{RGB}{190, 230, 60}  % #BEE63C
\definecolor{anisblue}{RGB}{60, 150, 250}   % #3C96FA
\definecolor{myolive}{RGB}{140,150,80}
\definecolor{myslate}{RGB}{100,120,160}

\usepackage[version=4]{mhchem}
\usepackage{chemfig}
\usepackage{siunitx}
\DeclareSIUnit{\atomic}{a.u.}
\usepackage{array}
\newcolumntype{M}[1]{>{\centering\arraybackslash}m{#1}}
\usepackage{paralist}
\usepackage{enumitem}
\usepackage{tikz}
\usetikzlibrary{arrows.meta,bending,calc,decorations.markings,patterns,shapes.symbols}
\usepackage{pgfplots}
\pgfplotsset{
  compat=newest,
  scale only axis,
  /pgf/number format/1000 sep={},
  every axis/.append style={
    scale only axis,
    grid style={line width=0.2pt}, % thinner, lighter gray
    major grid style={draw=gray!50},  % (optional) tweak major grid separately
    minor grid style={draw=gray!15},  % (optional) tweak minor grid separately
  },
}

\usepackage{natbib}
\def\vec#1{{{\underline{#1}}}}
\def\sio2{{\texorpdfstring{SiO$_2$}{}}}

\DeclareMathAlphabet\mathbfcal{OMS}{cmsy}{b}{n}
\DeclareMathOperator{\vectspan}{span}
\def\d{{\rm d}}

\def\etal{\emph{et al}}

\def\tang{\ThisStyle{\abovebaseline[0pt]{\scalebox{-1}{$\SavedStyle\perp$}}}}

\begin{document}

\title{The vibrational spectrum of vitreous silica:\\rigorous decomposition via recursive orthogonal splitting analysis}
%\title{Recursive orthogonal splitting analysis: a general approach to decompose the vibrational spectrum of disordered solids, illustrated by}
%\title{Hierarchical structure of the vibrational spectrum of vitreous silica\\ revealed by recursive orthogonal splitting}
%rigorous decomposition based on the projection formalism}

\author{Nikita S. Shcheblanov}
\affiliation{MSME, Univ Gustave Eiffel, Univ Paris Est Cr{\'e}teil, CNRS, 77454 Marne-la-Vall{\'e}e, France}
\affiliation{Navier, CNRS, Univ Gustave Eiffel, ENPC, Institut Polytechnique de Paris, Marne-la-Vallée, France}
\affiliation{Maastricht University, P.O. Box 616, 6200 MD Maastricht, The Netherlands}

%\author{Mikhail E. Povarnitsyn}
%\affiliation{Navier, Ecole des Ponts, Univ Gustave Eiffel, CNRS, 77420 Marne-la-Vallée, France}

\author{Anaël Lemaître}
\email{anael.lemaitre@enpc.fr}
\affiliation{Navier, CNRS, Univ Gustave Eiffel, ENPC, Institut Polytechnique de Paris, Marne-la-Vallée, France}

\date{\today}

\begin{abstract}
Our understanding of vibrations in solids currently rests on concepts and techniques designed for crystals and explicitly relying on periodicity, hence inapplicable to amorphous materials. As a consequence, no established framework enables a systematic decomposition of the vibrational spectrum of amorphous solids into contributions associated with well-defined types of atomic motions. This methodological gap obscures the interpretation of various experimental probes of linear response, based on the measurements of acoustic, thermal, or optical properties. Here, we construct such a framework---Recursive Orthogonal Splitting Analysis (ROSA)---which decomposes the vibrational space by recursive applications of the projection formalism. Each step of the procedure exploits a dominant stiffness contrast to split a vibrational displacement subspace into two weakly coupled orthogonal complements. We illustrate ROSA by applying it to the archetypal covalent glass---vitreous silica. Successively separating bond-stretching, symmetric and antisymmetric oxygen motions, isotropic and deviatoric tetrahedral strains, and distinct classes of tetrahedral bending, reveals a hierarchical structure of the space of vibrational degrees of freedom, involving six mutually orthogonal subspaces. These subspaces selectively capture all salient spectral features, including the two-humped structure in the low-frequency range, the peak near $\SI{800}{cm^{-1}}$, and the high-frequency doublet. In the low-frequency range, rigid-unit tetrahedral rotations do not constitute independent degrees of freedom but are kinematically enslaved to bending coordinates by no-stretch constraints. Because ROSA relies solely on the existence of contrasted stiffness scales associated with the point symmetry of local structural units, and on their separation by enforcing geometric constraints via the projection formalism, it is directly applicable to a broad class of covalent network glasses.
\end{abstract}

\maketitle

\section{Introduction}
The vibrational response of elastic solids governs a wide range of physical properties, from heat transport to acoustic signatures and optical activity. Understanding how it is determined by the microstructure is therefore essential both when constructing predictive theories of these properties and when using their experimental measurements as microstructural probes. The possibility of implementing these strategies in \emph{crystalline} solids has been foundational to the development of many technologies---from semiconductors to optical, thermoelectric, and superconducting devices~\cite{nikogosyan97,nikogosyan2005,pop2006,yariv2002,cahill2003,yu2010,cahill2014,snyder2008,giustino2017,simoncelli2019,qian2021,chen2022}---throughout the last century. This success, however, critically depends on simplifications afforded by periodic order and local symmetries~\cite{maradudin71,lax1974,bottger1983,birman1984}, which allow peaks and bands of the vibrational density of states (VDOS) to be interpreted in terms of distinct families of atomic motions.

No such framework exists for \emph{amorphous} solids, as the absence of periodicity precludes applying techniques designed for crystalline solids~\cite{elliott74}. Consequently, despite decades of effort~\cite{dean72,bell1972,sen77,laughlin77,laughlin78,galeener79,weaire1980,phillips83,wilson96,pavlatou97,taraskin97,sarnthein97,pasquarello98a,pasquarello98b,umari2005,pietrucci2008,mazzarello2010,povarnitsyn2020,shcheblanov2020,giacomazzi2023}, there is no established method to reliably assign features of the VDOS to well-defined families of atomic displacements. This fundamental gap obscures the interpretation of experimental probes of the linear vibrational response---from Raman and infrared spectroscopy to inelastic scattering techniques---and hampers our understanding of key anomalies arising from structural disorder such as the Boson peak~\cite{nakayama2002} and the low-temperature plateau in thermal conductivity~\cite{pohl2002}.

Vitreous silica provides an ideal test case for such an endeavor. Its structural organization into \ce{SiO4} tetrahedra connected by bridging oxygen atoms gives rise to only a few salient structural motifs~\cite{hanna65,bock70,sen77,galeener79}---such as tetrahedra or \ce{Si-O-Si} bridges---which exhibit a pronounced local anisotropy and thus introduce a systematic stiffness contrast between different types of small-scale atomic motion. This feature is expected to play a central role in determining the relationship between microstructure and vibrational properties. Moreover, its VDOS exhibits prominent spectral features---including a two-humped structure in the low-frequency band range, a peak around \SI{800}{cm^{-1}}, and a high-frequency doublet~\cite{leadbetter72,price87,sarnthein97,pasquarello98a,giacomazzi09}---that are widely believed to originate from specific families of atomic motions~\cite{dean72,laughlin77,sen77,galeener79,weaire1980,phillips83,galeener83,mcmillan84,wilson96,taraskin97,sarnthein97,pasquarello98a}.

Numerous studies have sought to rationalize the vibrational spectrum of silica by decomposing the associated eigenmodes into various local contributions. The simplest approach, pioneered by Bell and Dean~\cite{bell1972,dean72}, separates the contributions of oxygen and silicon atoms, and decomposes the motion of each oxygen atom in a basis defined by its two neighboring silicon atoms~\cite{bell68,bell70a,bell70b,bell71,bell1972,dean72,bell75,taraskin97,pasquarello98a,pasquarello98b}. However, this decomposition offers only limited insight: it fails to establish a clear correspondence between these atomic degrees of freedom and distinct features in the VDOS~\cite{wilson96,taraskin97}.

The problem evidently stems from the strong covalent bonding between oxygen and silicon atoms, leading researchers to analyze the vibrational problem in terms of nested, bonded, molecular-like structural units such as \ce{SiO4} tetrahedra or \ce{Si-O-Si} bridging groups~\cite{laughlin77,laughlin78,mcmillan84,guillot97a,guillot97b,guimbretiere2008,hehlen12}. The point symmetry groups of these units~\cite{decius1977,hamermesh1962,herzberg1991}---namely $T_d$ for \ce{SiO4} tetrahedra and $C_{2v}$ for \ce{Si-O-Si} bridges---provide a natural framework for classifying local deformations into symmetry-adapted modes that typically exhibit distinct stiffnesses. This idea motivated the development of the \emph{local projection approach}~\cite{wilson96,pavlatou97,guillot97a,guillot97b,sarnthein97,pasquarello98a,pasquarello98b,taraskin97,shcheblanov16,umari2005}, in which vibrational eigenmodes are projected onto symmetry-adapted local modes at the level of structural units. Analyses based on this framework reveal that distinct local tetrahedral motions---such as rotations, stretch and bending deformations---contribute preferentially to different frequency ranges of the vibrational spectrum in vitreous silica.

However, the local projection approach lacks a rigorous mathematical foundation: it does not perform a true (global) projection of vibrational eigenmodes onto mutually orthogonal subspaces, and therefore cannot ensure strict orthogonality between the different components it considers, leading to overcompleteness issues. As a result, it does not allow for a definitive assignment of vibrational modes, that is, an unambiguous identification of the specific classes of atomic motion underlying distinctive features of the vibrational spectrum.

Here, we introduce a general framework of analysis of the vibrational spectrum of covalent glasses, which we call recursive orthogonal splitting analysis (ROSA), and we show how it can be used to decompose the vibrational spectrum of vitreous silica. ROSA consists in performing recursive orthogonal splittings of the vibrational degrees of freedom using the projection formalism~\cite{cl2020,meyer2023}. Each step of the procedure exploits a dominant stiffness contrast to split the vibrational problem into two weakly coupled components associated with distinct frequency ranges. Successive splittings then enable the decomposition of the full vibrational problem into a set of weakly coupled, mutually orthogonal complements, each characterized by its intrinsic vibrational frequencies and dominating a specific frequency range.

Applying ROSA to a model of vitreous silica generated using a combination of classical and \emph{ab initio} simulations---thereby providing first-principles interatomic interactions and Hessian matrix---we uncover a hierarchical structure of the vibrational space, which directly reflects the stiffness contrasts between distinct classes of atomic motions of well-defined dimensionality. Specifically, we show that in an $N$-atom silica system having $3N$ degrees of freedom, the lower part of the spectrum is dominated by a \emph{no-stretch} subspace of dimension $5N/3$, consisting of displacement fields that preserve all \ce{Si-O} bond lengths. The orthogonal complement, a \emph{bond-stretch} eigenspace, of dimension $4N/3$, can be constructed explicitly as the linear combinations of contra-directional (out-of-phase) displacements of bonded silicon and oxygen atoms.

The low-frequency no-stretch subspace can be further decomposed on the basis of a stiffness contrast between tetrahedral bending degrees of freedom. Namely, a low-frequency $2N/3$-dimensional subspace is identified by isolating modes in which tetrahedral distortions exclusively consist of \emph{symmetric bending} (or \emph{torsion}) ($E$-type), while excluding \emph{antisymmetric bending}. Its $N$-dimensional complement accounts for the upper part of the no-stretch subspace and the peak near \SI{400}{cm^{-1}}.

Within the high-frequency bond-stretch subspace, the dominant stiffness contrast originates from the $2N/3$ elementary motions in which oxygen atoms move along the axis connecting the two silicon atoms they are bonded to. When the associated contra-directional motion of silicon is properly taken into account, this defines an \emph{antisymmetric-stretch} subspace exactly corresponding to the high-frequency doublet of the VDOS, while its $2N/3$-dimensional complement gives rise to a mid-frequency band. Each of these two subspaces can be further decomposed into two $N/3$-dimensional subspaces based on the stiffness contrast between \emph{isotropic} and \emph{deviatoric} tetrahedral stretches.

Through this recursive process, we identify six elementary, mutually orthogonal subspaces, each associated with distinct, albeit partially overlapping, frequency domains that account for the main features of the VDOS in vitreous silica.\\

The article is organized as follows. In Section~\ref{sec:projection}, we introduce the vibrational problem and present the projection formalism together with the mathematical tools necessary to implement an orthogonal decomposition of the vibrational dynamics. In Section~\ref{sec:model}, we describe the numerical model of vitreous silica, present its VDOS, and illustrate state-of-the-art decompositions of eigenvectors in terms of atomic contributions and via the local projection analysis. In Section~\ref{sec:decomposition}, we implement ROSA to decompose the vibrational spectrum of vitreous silica: first, we split the full polarization subspace according to \ce{Si-O} bond stretching; then, we analyze the bond-stretch subspace on the basis of oxygen motion and then tetrahedral stretches; finally, we turn to the low-frequency no-stretch subspace.

\section{Vibrational spectrum, projection formalism, and local symmetries}
\label{sec:projection}

\subsection{Vibrational problem and density of states}
The vibrational dynamics of an $N$-atom system around a local minimum of the potential energy landscape is governed by the linearized equation of motion for the $3N$-dimensional vector $\mathbf{u}\equiv\{\vec u_i\}$ of atomic displacements~\cite{maradudin71}:
\begin{equation}\label{eq:vib}
  \mathbfcal{M}\,\mathbf{\ddot u}=-\mathbfcal{H}\,\mathbf{u} \, ,
\end{equation}
where $\mathbfcal{H}$ denotes the Hessian matrix, which is real-valued, symmetric, and positive-semidefinite, and $\mathbfcal{M}$ denotes the mass matrix, which is block-diagonal with blocks $m_i\mathbf{I}_3$ for $i=1,\ldots, N$, with $\mathbf{I}_3$ being the $3\times3$ identity matrix.

This problem is classically solved by introducing the mass-weighted vibrational displacement vector $\mathbf{e}=\mathbfcal{M}^{1/2}\,\mathbf{u}$ and the dynamical matrix $\mathbfcal{D}=\mathbfcal{M}^{-1/2}\mathbfcal{H}\mathbfcal{M}^{-1/2}$, which transforms Eq.~\eqref{eq:vib} into:
\begin{equation}\label{eq:dynamics}
  \mathbf{\ddot e}=-\mathbfcal{D}\,\mathbf{e}\,.
\end{equation}
The normal modes of vibration are solutions of the form $\mathbf{e}_p\,e^{i\omega_p t}$ of this equation, where the polarization vectors $\mathbf{e}_p$ and the eigenfrequencies $\omega_p$ satisfy the eigenvalue problem:
\begin{equation}
  \mathbfcal{D}\,\mathbf{e}_p=\omega_p^2\,\mathbf{e}_p\,.
\end{equation}
Since the dynamical matrix $\mathbfcal{D}$ is real, symmetric, and positive-semidefinite, its eigenvectors form a complete orthonormal basis of the $3N$-dimensional vector space of polarization vectors denoted $\mathbf{E}$.
The VDOS is defined as the distribution of eigenfrequencies:
\begin{equation}\label{eq:VDOS}
\rho(\omega)=\frac{1}{3N-3}\,\sum_p\,\delta(\omega-\omega_p) \,,
\end{equation}
where the sum runs over all modes, and the normalization factor takes into account the existence of 3 zero-frequency modes resulting from translation invariance.

\subsection{Orthogonal decomposition}
An orthogonal decomposition of $\mathbf{E}$, denoted $\mathbf{E}=\mathbf{E}_\mathrm{A}\oplus \mathbf{E}_\mathrm{A}^\perp$, is defined by a pair of orthogonal subspaces $\mathbf{E}_\mathrm{A}$ and $\mathbf{E}_\mathrm{A}^\perp$ that generate $\mathbf{E}$, i.e., such that every vector $\mathbf{e}\in\mathbf{E}$ can then be written as $\mathbf{e}=\mathbf{e}_\mathrm{A}+\mathbf{e}_\mathrm{A}^\perp$ with $\mathbf{e}_\mathrm{A}\in\mathbf{E}_\mathrm{A}$ and $\mathbf{e}_\mathrm{A}^\perp\in\mathbf{E}_\mathrm{A}^\perp$. As a consequence of orthogonality, this decomposition is unique and $\mathbf{e}_\mathrm{A}\cdot\mathbf{e}_\mathrm{A}^\perp=0$. The considered subspaces are then said to be orthogonal complements. The orthogonal projections onto $\mathbf{E}_\mathrm{A}$ and $\mathbf{E}_\mathrm{A}^\perp$ are symmetric matrices denoted $\mathbfcal{P}_\mathrm{A}$ and $\mathbfcal{P}_\mathrm{A}^\perp$, respectively. They satisfy $\mathbfcal{P}_\mathrm{A}^\perp+\mathbfcal{P}_\mathrm{A}=\mathbfcal{I}$ with $\mathbfcal{I}$ being the identity operator on $\mathbf{E}$.

The partial VDOS associated with any subspace $\mathbf{E}_\mathrm{A}\subset\mathbf{E}$ is defined as:
\begin{equation}
\label{eq:partial}
  \varrho_\mathrm{A}(\omega)=\frac{1}{3N-3}\,\sum_p\,\|\mathbfcal{P}_\mathrm{A}\mathbf{e}_p\|^2 \delta(\omega-\omega_p)\,,
\end{equation}
with $\mathbf{e}_p$ the $p$-th eigenmode. Provided an orthogonal decomposition $\mathbf{E}=\mathbf{E}_\mathrm{A}\oplus \mathbf{E}_\mathrm{A}^\perp$, the full VDOS decomposes as: $\rho=\varrho_\mathrm{A}+\varrho_\mathrm{A}^\perp$. This property follows solely from the orthogonality of $\mathbf{E}_\mathrm{A}$ and $\mathbf{E}_\mathrm{A}^\perp$, which ensures that, for any normalized eigenmode, $\mathbf{e}_p$, $\|\mathbfcal{P}_\mathrm{A}\mathbf{e}_p\|^2+\|\mathbfcal{P}_\mathrm{A}^\perp\mathbf{e}_p\|^2=1$. It does not require $\mathbf{E}_\mathrm{A}$ and $\mathbf{E}_\mathrm{A}^\perp$ to coincide with vibrational eigenspaces. The partial VDOS $\varrho_\mathrm{A}$ therefore quantifies the contribution of subspace $\mathbf{E}_\mathrm{A}$ to vibrational modes at a given frequency.

\subsection{Projection formalism}
In the context of vibrational dynamics, the projection formalism provides a framework to decompose the eigenvalue problem $\omega^2\,\mathbf{e}=\mathbfcal{D}\,\mathbf{e}$ onto complementary subspaces of the polarization space~\cite{cl2020}. This decomposition can be performed recursively and hierarchically to isolate increasingly specific groups of atomic motion. But the underlying principles are captured by considering an orthogonal splitting of the polarization space, $\mathbf{E}=\mathbf{E}_\mathrm{A}\oplus \mathbf{E}_\mathrm{A}^\perp$ and the corresponding decomposition of the eigenvalue problem as:
\begin{equation}
    \left\{
    \begin{aligned}
    \omega^2\,\mathbf{e}_\mathrm{A}&=\mathbfcal{D}_\mathrm{A}^{\tang\tang}\,\mathbf{e}_\mathrm{A}+\mathbfcal{D}_\mathrm{A}^{\tang\perp}\,\mathbf{e}_\mathrm{A}^\perp\\
    \omega^2\,\mathbf{e}_\mathrm{A}^\perp&=\mathbfcal{D}_\mathrm{A}^{\perp\tang}\,\mathbf{e}_\mathrm{A}+\mathbfcal{D}_\mathrm{A}^{\perp\perp}\,\mathbf{e}_\mathrm{A}^\perp
    \end{aligned}
    \right. \,,
    \label{eq:decomposition}
\end{equation}
where the projected components of the mass-weighted displacements are defined as $\mathbf{e}_\mathrm{A}=\mathbfcal{P}_\mathrm{A}\,\mathbf{e}$, $\mathbf{e}_\mathrm{A}^\perp=\mathbfcal{P}_\mathrm{A}^\perp\,\mathbf{e}$, and the projected operators are
\begin{equation}
\begin{aligned}
&\mathbfcal{D}_\mathrm{A}^{\tang\tang}=\mathbfcal{P}_\mathrm{A}\mathbfcal{D}\mathbfcal{P}_\mathrm{A} \,, \\
&\mathbfcal{D}_\mathrm{A}^{\tang\perp}=\mathbfcal{P}_\mathrm{A}\mathbfcal{D}\mathbfcal{P}_\mathrm{A}^\perp \,, \\
&\mathbfcal{D}_\mathrm{A}^{\perp\tang}=\mathbfcal{P}_\mathrm{A}^\perp\mathbfcal{D}\mathbfcal{P}_\mathrm{A} \,, \\
&\mathbfcal{D}_\mathrm{A}^{\perp\perp}=\mathbfcal{P}_\mathrm{A}^\perp\mathbfcal{D}\mathbfcal{P}_\mathrm{A}^\perp \,.
\end{aligned}
\end{equation}
Equation~\eqref{eq:decomposition} is an exact reformulation of the full eigenvalue problem, in which the dynamical matrix $\mathbfcal{D}$ is represented in block form.

The interest of the procedure lies in the fact that the diagonal blocks account for intrinsic vibrational properties of the individual subspaces, which can be probed by considering the \emph{restricted} eigenvalue problems
\begin{equation}\label{eq:A}
  \omega_\mathrm{A}^2\,\mathbf{e}_\mathrm{A}=\mathbfcal{D}_\mathrm{A}^{\tang\tang}\,\mathbf{e}_\mathrm{A}
\end{equation}
on $\mathbf{E}_\mathrm{A}$ and
\begin{equation}\label{eq:Aperp}
  \left(\omega_\mathrm{A}^\perp\right)^2\,\mathbf{e}_\mathrm{A}^\perp=\mathbfcal{D}_\mathrm{A}^{\perp\perp}\,\mathbf{e}_\mathrm{A}^\perp
\end{equation}
on $\mathbf{E}_\mathrm{A}^\perp$. Taken together, Eqs.~\eqref{eq:A} and~\eqref{eq:Aperp} define an \emph{uncoupled} vibrational problem for which $\mathbf{E}_\mathrm{A}$ and $\mathbf{E}_\mathrm{A}^\perp$ are, by construction, eigenspaces. The VDOS of this uncoupled problem is given by the sum $\rho_\mathrm{A}+\rho_\mathrm{A}^\perp$, where $\rho_\mathrm{A}$ and $\rho_\mathrm{A}^\perp$ denote the VDOS of the restricted problems~\eqref{eq:A} and~\eqref{eq:Aperp}, respectively.

From this perspective, Eq.~\eqref{eq:decomposition} represents the whole vibrational problem as resulting from the coupling between the $\mathbf{E}_\mathrm{A}$ and $\mathbf{E}_\mathrm{A}^\perp$ subsystems through the off-diagonal block operators. This decomposition is, of course, most informative when these off-diagonal coupling operators introduce only limited mixing between the subspaces, so that $\mathbf{E}_\mathrm{A}$ and $\mathbf{E}_\mathrm{A}^\perp$ capture distinct features of the full vibrational spectrum $\rho$.

\subsection{Constructing projectors}
\label{sec:projectors}
A fundamental result of linear algebra states that, for any linear operator $\mathbfcal{A}$ with image in $\mathbf{E}$, the image subspace $\mathbf{E}_\mathrm{A}=\mathrm{Im}(\mathbfcal{A})$ and the co-kernel $\mathbf{E}_\mathrm{A}^\perp=\mathrm{Ker}(\mathbfcal{A}^\intercal)$ form orthogonal complementary subspaces of $\mathbf{E}$. Conversely, any pair of orthogonal subspaces can be represented in this form for a suitable choice of a linear operator (or matrix) $\mathbfcal{A}$. This result is fully general and does not require $\mathbfcal{A}$ to have full column rank.

In this construction, let $\mathbfcal{A}$ be an $m\times n$ matrix, with $m=\dim\mathbf{E}$, mapping an $n$-component coordinate vector $\boldsymbol{\alpha}$ onto elements of the polarization space $\mathbf{E}$: $\mathbf{e}=\mathbfcal{A}\,\boldsymbol{\alpha}\in\mathbf{E}$. The image subspace $\mathbf{E}_\mathrm{A}=\mathrm{Im}(\mathbfcal{A})\subset\mathbf{E}$ is generated by the column vectors of $\mathbfcal{A}$, while the co-kernel $\mathbf{E}_\mathrm{A}^\perp=\mathrm{Ker}(\mathbfcal{A}^\intercal)  $ consists of all vectors $\mathbf{e}\in\mathbf{E}$ satisfying the orthogonality constraints $\mathbfcal{A}^\intercal\mathbf{e}=\mathbf{0}$. In the following, this representation will provide a convenient and general framework to construct orthogonal decompositions of the polarization space.

The orthogonal projection onto $\mathbf{E}_\mathrm{A}=\mathrm{Im}(\mathbfcal{A})$ is then given by $\mathbfcal{P}_\mathrm{A}=\mathbfcal{A}\mathbfcal{A}^+$, where $\mathbfcal{A}^+$ is the Moore-Penrose pseudoinverse of $\mathbfcal{A}$, an $n\times m$ matrix the calculation of which is described in Appendix~\ref{app:projectors}. The projector onto the complementary subspace $\mathbf{E}_\mathrm{A}^\perp$ is then $\mathbfcal{P}_\mathrm{A}^\perp=\mathbfcal{I}-\mathbfcal{P}_\mathrm{A}$.

When decomposing subspaces recursively, we will encounter situations where it is necessary to determine the orthogonal projector onto the intersection $\mathbf{G}\cap\mathbf{H}$ of two subspaces $\mathbf{G}$ and $\mathbf{H}$, given their associated orthogonal projectors $\mathbfcal{P}_\mathrm{G}$ and $\mathbfcal{P}_\mathrm{H}$. The orthogonal projector onto the intersection subspace is denoted $\mathbfcal{P}_\mathrm{G}\wedge\mathbfcal{P}_\mathrm{H}$. The wedge operator is commutative, $\mathbfcal{P}_\mathrm{G}\wedge\mathbfcal{P}_\mathrm{H}=\mathbfcal{P}_\mathrm{H}\wedge\mathbfcal{P}_\mathrm{G}$, and reduces to the product $\mathbfcal{P}_\mathrm{G}\wedge\mathbfcal{P}_\mathrm{H}=\mathbfcal{P}_\mathrm{G}\mathbfcal{P}_\mathrm{H}=\mathbfcal{P}_\mathrm{H}\mathbfcal{P}_\mathrm{G}$ whenever the two projectors commute (see Appendix~\ref{app:projectors}).

\subsection{Point symmetries and symmetry-adapted modes of local structural units}

Our analysis will rely on basic notions from group theory, which explain how the point symmetry of a regular structural unit endows the space of its local deformations with an intrinsic structure organized into classes of deformation modes that are equivalent under symmetry operations and hence correspond to distinct stiffness scales.

Regular structural units (like molecules) possess a point symmetry group $G$, defined as the set of spatial operations---rotations and reflections---that leave its reference configuration invariant~\cite{decius1977}. In the case of vitreous silica, relevant groups are $C_{2v}$ for \ce{Si-O-Si} bridges and $T_d$ for \ce{SiO4} tetrahedra.

Each symmetry operation $g\in G$ acts naturally on the displacement fields of the considered unit: it permutes atomic labels and transforms the displacement vectors according to the corresponding spatial transformation. Hence, every $g$ defines a linear transformation $D(g)$ on the space of local displacements. The correspondence $g\to D(g)$ is called a \emph{representation} of the group $G$.

The space of local displacements thus carries a representation of $G$. Any subspace which is invariant under all transformations $D(g)$ (with $g\in G$) also carries a representation, obtained by restricting $D(g)$ on that subspace. A representation is said to be \emph{irreducible} when it admits no non-trivial $G$-invariant subspace.

A \emph{symmetry-adapted} basis is composed of vectors belonging to irreducible subspaces. In such a basis, any linear transformation $D(g)$ takes a block-diagonal form, in which each block corresponds to an irreducible subspace.

Group theory ensures that every finite group possesses a well-defined set of irreducible representations. Since the space of local displacements carries a representation of the point symmetry group, it decomposes uniquely into invariant subspaces, each carrying an irreducible representation. These irreducible subspaces are minimal non-trivial displacement subspaces that remain invariant under the symmetry operations.

Since the internal energy of the structural unit is invariant under point symmetry operations, its local Hessian commutes with the action of the group. Each irreducible subspace is therefore an eigenspace of the local Hessian and can be associated with a distinct stiffness. Accordingly, symmetry-adapted modes, the basis vectors spanning irreducible subspaces, provide a natural framework to separate local stiffness contributions.

\section{Silica model and state-of-the-art analyses of atomic vibrations}
\label{sec:model}
In this section, we introduce the atomistic model of vitreous silica used throughout this study and analyze its VDOS using existing approaches---namely, decomposition into atomic degrees of freedom and local projection analysis---in order to provide baseline data representative of the current state of the art. The implementation of ROSA for this silica model is presented in Section~\ref{sec:decomposition}.

\subsection{Model preparation, structural, and vibrational properties}
In order to obtain a reliable representation of the VDOS, we use a 648-atom structural model of vitreous silica prepared by combining classical molecular dynamics (MD) and \emph{ab initio} simulations.

To rapidly generate a structurally relaxed configuration, we first performed classical MD simulations using LAMMPS~\cite{lammps} and the van Beest-Kramer-van Santen (BKS) interatomic potential~\cite{bks1990}, with Wolf's truncation of Coulombic interactions~\cite{carre2007,shcheblanov2015,shcheblanov2020}. Within the NPT ensemble at ambient pressure, a 648-atom supercell of $\beta$-cristobalite was heated to \SI{5200}{K}, equilibrated at this temperature for \SI{1}{\nano\second}, and subsequently quenched to \SI{1}{K} at a rate of \SI{1e11}{K/s}. The resulting configuration is amorphous and has a density of approximately $\SI{2.2}{g/cm^3}$, consistent with that of vitreous silica.

This configuration is then relaxed using the \texttt{QUICKSTEP} method of the CP2K package~\cite{cp2k} to minimize the total energy and access the corresponding \emph{inherent state} within a first-principles framework. Specifically, we perform DFT calculations in the local density approximation (LDA) using Gaussian based pseudopotentials with a TZVP atomic basis set from the CP2K library and expanding the electron density in plane waves with an energy cutoff of \SI{700}{Ry}. Within this framework, we minimize the total energy, using as  convergence criteria a maximum force threshold of \SI{1e-6}{Ha/bohr} and a maximum atomic displacement between successive iterations of $\lesssim\SI{1e-4}{bohr}$.

\begin{figure}[b]
  \includegraphics[width=0.47\textwidth]{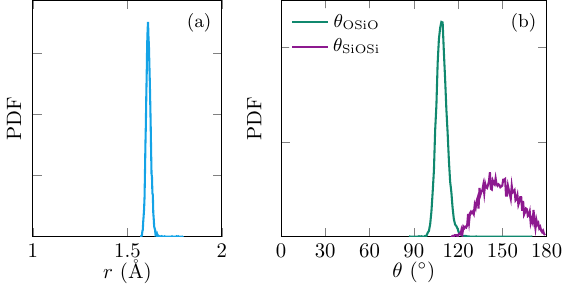}
  \caption{\label{fig:model} Normalized distributions: (a) \ce{Si-O} bond lengths; (b) \ce{O-Si-O} and \ce{Si-O-Si} angles, as averaged over 8 samples.}
\end{figure}

The microstructure is analyzed by defining covalent bonds as \ce{Si-O} pairs separated by distances below $\SI{2}{\angstrom}$. We thus identify 8 configurations that do not contain any coordination defect, i.e., in which the \ce{Si-O}-bond network consists exclusively of tetrahedrally coordinated silicon atoms connected through bridging oxygen atoms. All the numerical data presented in the article are obtained by averaging over these configurations. The basic structural characteristics of these samples are shown in Fig.~\ref{fig:model}. As expected~\cite{giacomazzi09,trease2017,srivastava2018}, the bond lengths and \ce{O-Si-O} angles are very narrowly distributed around \SI{1.6}{\angstrom} and \SI{109}{\degree}, respectively. The \ce{Si-O-Si} angles exhibit a broader distribution than the \ce{O-Si-O} angles but are systematically $\gtrsim2\pi/3$.

For each sample, the Hessian matrix is computed by finite difference as follows: (i) each atom is sequentially displaced by an increment $\pm\,\texttt{DX}$ along each Cartesian direction (with $\texttt{DX}=\SI{5e-3}{\atomic}$); (ii) for each such perturbation, the forces acting on all atoms are calculated using a self-consistent field (\texttt{SCF}) convergence criterion of $\texttt{EPS\_SCF}=10^{-8}$; and (iii) a three-point finite-difference scheme is applied to obtain the linear response.

\begin{figure}[t]
  \includegraphics[width=0.47\textwidth]{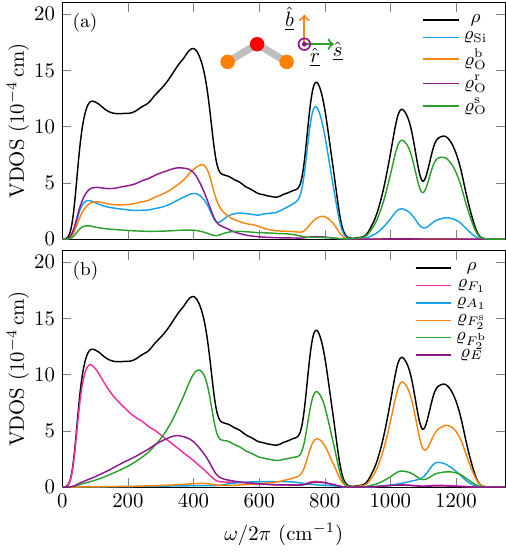}
  \caption{\label{fig:local} Total VDOS of the silica model ($\rho$, black) and two types of eigenmode analyses. (a) Eigenvector decomposition on atomic degrees of freedom~\cite{dean72}, showing the partial densities $\varrho_{\ce{Si}}$, $\varrho_{\ce{O}}^{\mathrm{s}}$, $\varrho_{\ce{O}}^{\mathrm{b}}$, and $\varrho_{\ce{O}}^{\mathrm{r}}$, which capture the contributions of subspaces $\mathbf{E}_{\ce{Si}}$, $\mathbf{E}_{\ce{O}}^{\mathrm{s}}$, $\mathbf{E}_{\ce{O}}^{\mathrm{b}}$, and $\mathbf{E}_{\ce{O}}^{\mathrm{r}}$, respectively. The inset presents a sketch of the local stretching, bending, rocking basis associated with an \ce{Si-O-Si} triplet, with silicon atoms shown in orange and oxygen in red. (b) Local projection analysis including tetrahedral rotations and shape degrees of freedom, following Refs.~\cite{taraskin97,shcheblanov16}.}
\end{figure}

Since we deal with small systems, the VDOS and partial VDOS are calculated, as is common, by replacing the Dirac delta distribution in Eqs.~\eqref{eq:VDOS} and~\eqref{eq:partial} with a Gaussian function, which we will take of width of \SI{10}{cm^{-1}} throughout the text.

The VDOS thus calculated from our \emph{ab initio} silica model is shown in Fig.~\ref{fig:local}. It exhibits salient features such as a double-humped band-like structure in the low-frequency range, a sharp peak near \SI{800}{cm^{-1}}, and a high-frequency doublet. When data are displayed over the full frequency range---as is done here and throughout the paper---sample-to-sample fluctuations are barely discernible in the total VDOS, as well as in the partial and restricted VDOS spectra presented in subsequent analyses.

\subsection{State-of-the-art vibrational analyses}
\label{sec:art}
Here, we illustrate prior attempts to assign VDOS features to families of atomic displacements, either through the separation of different atomic degrees of freedom~\cite{bell68,bell70a,bell70b,bell71,bell1972,dean72,bell75,taraskin97,pasquarello98a,pasquarello98b}, or through the local projection approach~\cite{wilson96,taraskin97}.

\subsubsection{Eigenmode decomposition on atomic degrees of freedom}
The most direct way to decompose the vibrational spectrum consists in separating the degrees of freedom associated with silicon and oxygen atoms. Additionally, it has long been recognized that oxygen motion exhibits a systematic stiffness contrast when expressed in a local basis aligned with the corresponding \ce{Si-O-Si} triplet~\cite{dean72}. This local basis defines three orthogonal directions as sketched in Fig.~\ref{fig:local}-(a): stretching ($\hat{\vec s}$, along the \ce{Si-Si} axis), bending ($\hat{\vec b}$, perpendicular to stretching in the \ce{Si-O-Si} plane), and rocking ($\hat{\vec r}$, perpendicular to the \ce{Si-O-Si} plane). This construction leads to the decomposition of the polarization space $\mathbf{E}=\mathbf{E}_{\ce{Si}}\oplus\mathbf{E}_{\ce{O}}^{\mathrm{b}}\oplus\mathbf{E}_{\ce{O}}^{\mathrm{r}}\oplus\mathbf{E}_{\ce{O}}^{\mathrm{s}}$, where $\mathbf{E}_{\ce{Si}}$ contains all displacement components associated with silicon atoms, and $\mathbf{E}_{\ce{O}}^{\mathrm{s}}$, $\mathbf{E}_{\ce{O}}^{\mathrm{b}}$, and $\mathbf{E}_{\ce{O}}^{\mathrm{r}}$ contain the stretching, bending, and rocking components of oxygen motion, respectively.

The partial VDOS spectra associated with these complementary subspaces are shown in Fig.~\ref{fig:local}-(a). This figure makes it clear that no feature of the VDOS is governed exclusively by a single atomic degree of freedom within this decomposition, as previously noted by Taraskin and Elliott in their analysis of VDOS spectra obtained from a classical MD model of silica~\cite{taraskin97}. The clearest separation is observed in the peak around \SI{800}{cm^{-1}}, which predominantly involves silicon motion, with a much weaker contribution from oxygen bending. The high-frequency doublet is primarily associated with oxygen stretching, but still exhibits a significant silicon contribution. By contrast, the low frequency region involves a complex combination of multiple atomic degrees of freedom. Overall, while certain trends can be identified, this decomposition does not reveal orthogonal subspaces that can be unambiguously associated with distinct spectral features.

\subsubsection{Local projection analysis}
\label{sec:local}

Another line of work~\cite{sen77,laughlin77,weaire1980,phillips83,wilson96,taraskin97,shcheblanov16}, inspired by the theory of molecular normal modes~\cite{decius1977,hamermesh1962,herzberg1991}, seeks to assess how the symmetry-adapted modes of elementary structural units such as \ce{Si-O-Si} triplets or \ce{SiO4} tetrahedra, contribute to the vibrational eigenvectors. This approach leads to the definition of a weighted VDOS of the form~\cite{taraskin97,shcheblanov16}:
\begin{equation}
\varrho_G(\omega)=\frac{1}{3N-3}\,\sum_p\,\left\langle\tilde r_{G,p}^2\right\rangle \delta(\omega-\omega_p) \,,
\end{equation}
where $\left\langle\tilde r_{G,p}^2\right\rangle$ denotes the average squared amplitude of the projection of eigenmode $p$ onto an irreducible subspace $G$ of the considered structural units~\cite{decius1977,hamermesh1962,herzberg1991}. Details about the procedure can be found in Refs.~\cite{taraskin97,shcheblanov16}.

\begin{figure}[t]
  \begin{center}
    \includegraphics[width=0.1\textwidth]{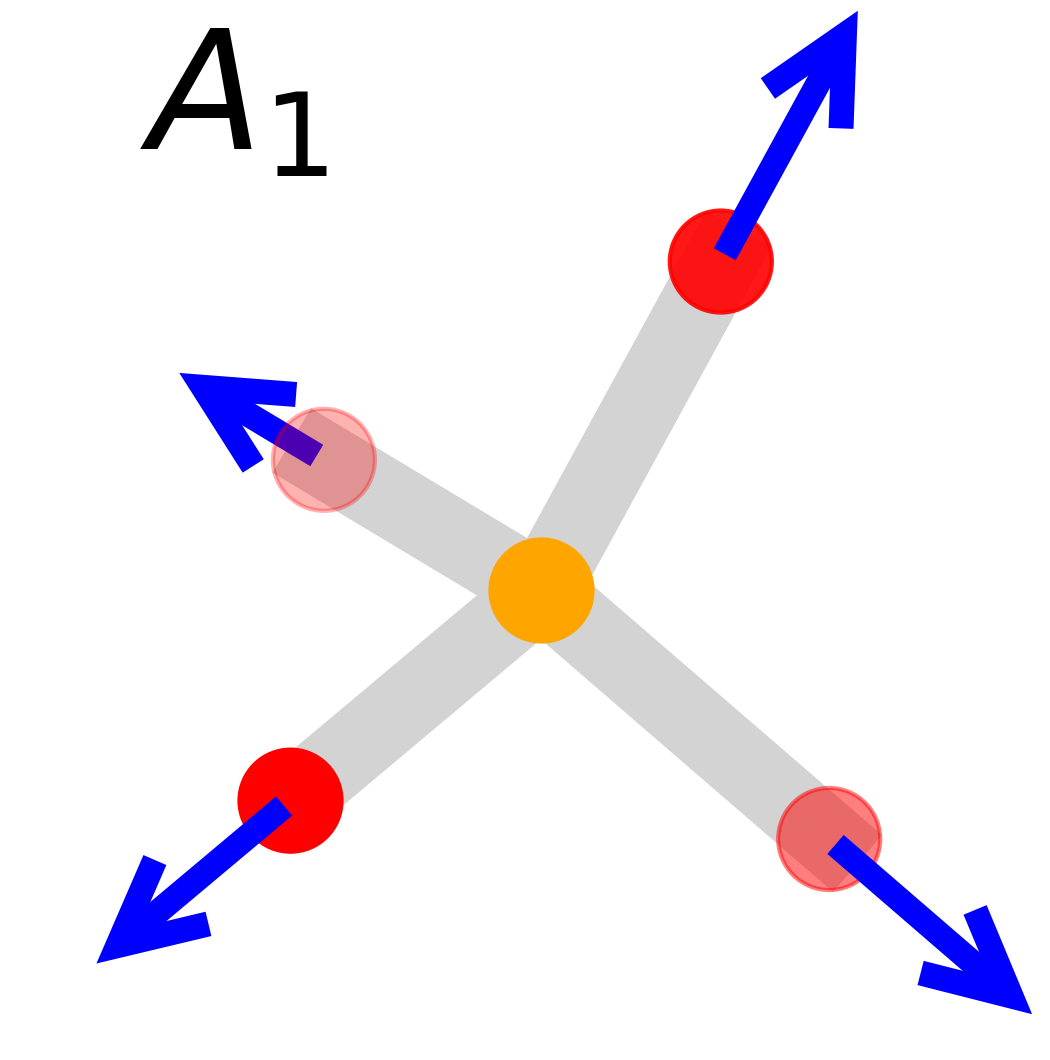}\hfil
    \includegraphics[width=0.1\textwidth]{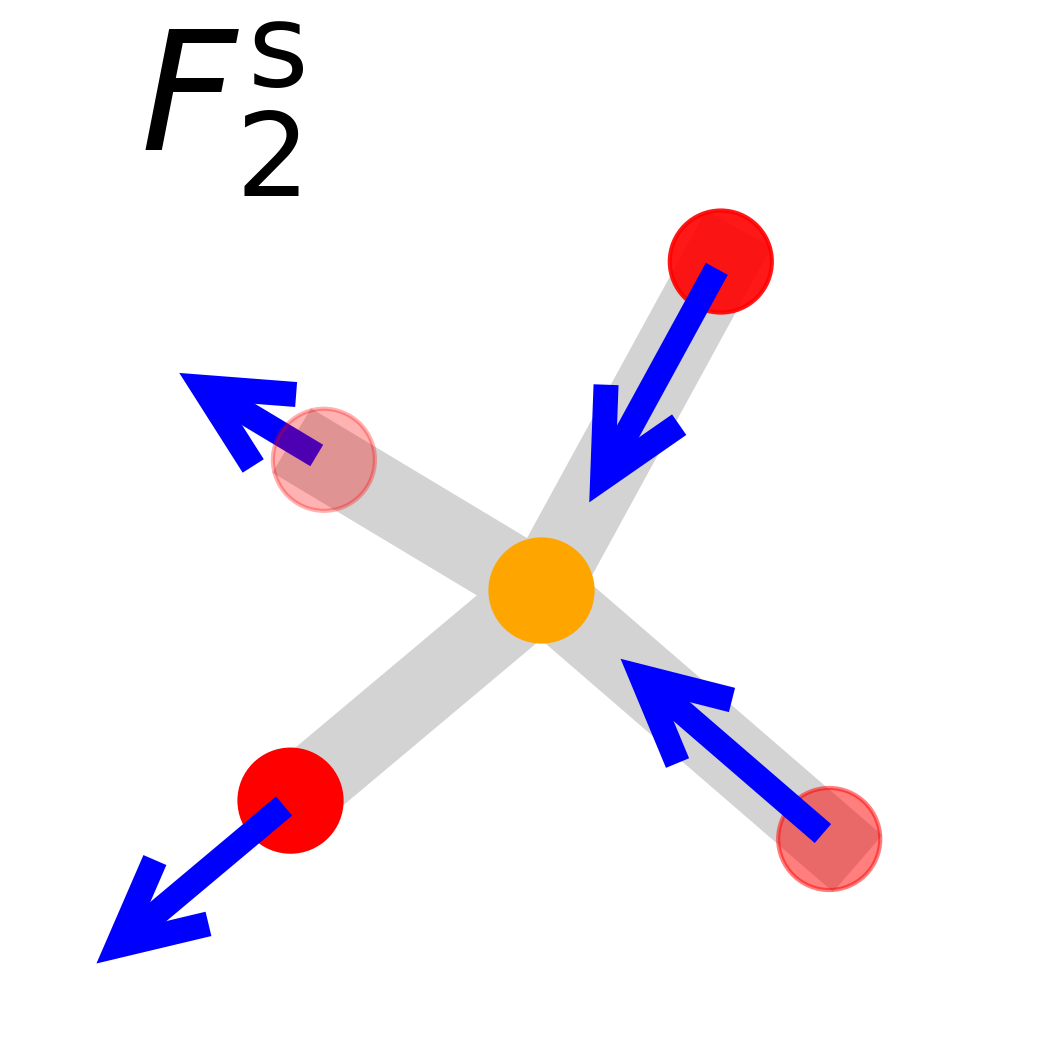}\hfil
    \includegraphics[width=0.1\textwidth]{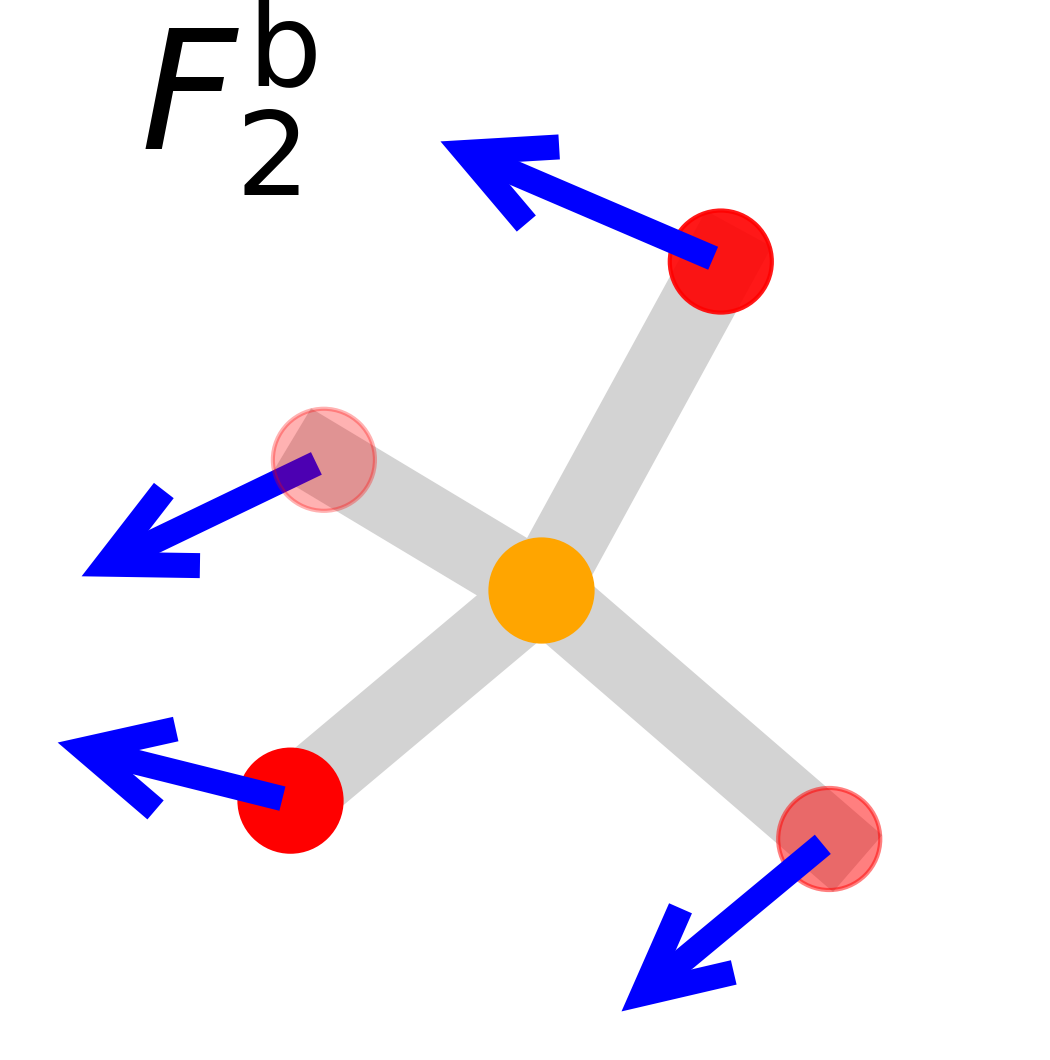}\hfil
    \includegraphics[width=0.1\textwidth]{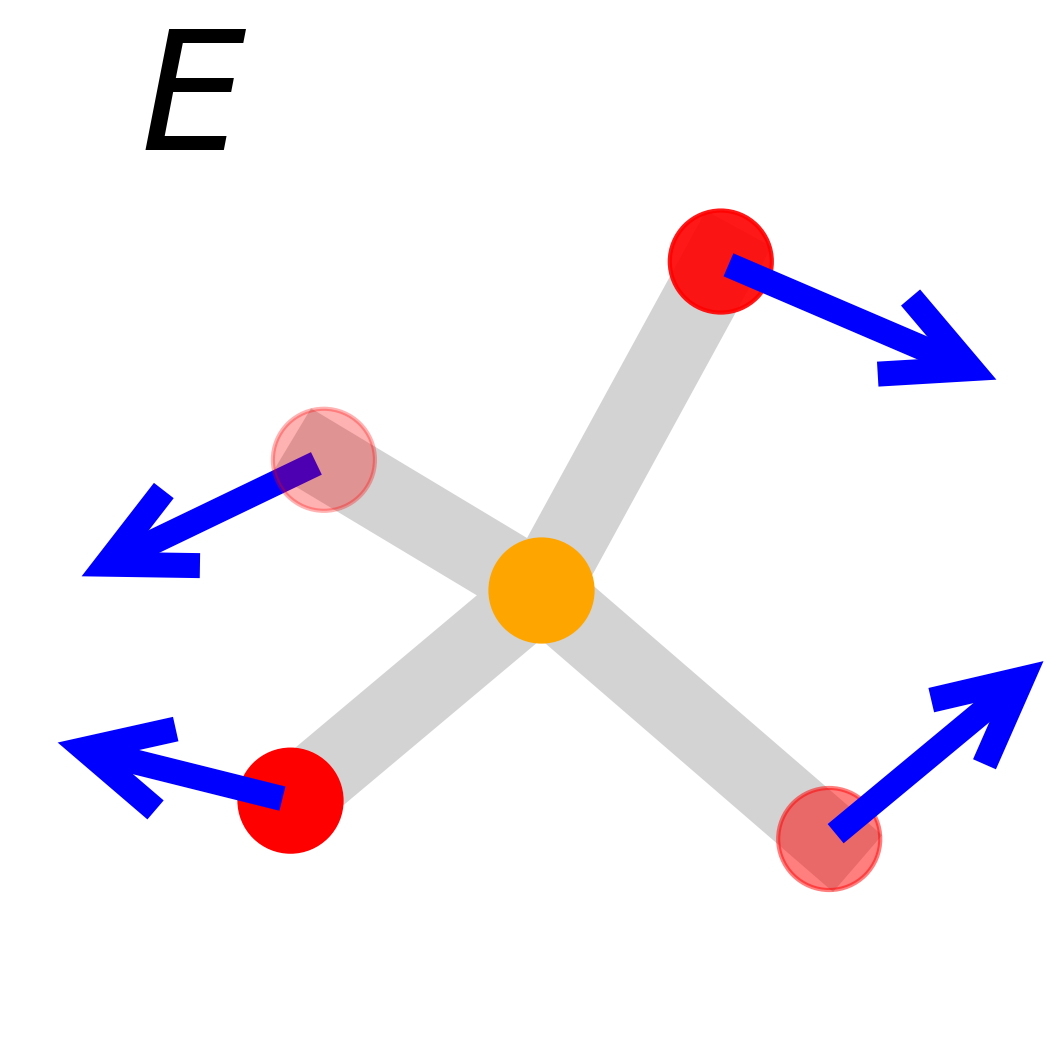}
  \end{center}
  \caption{\label{fig:tetra} Elementary displacements in a reference frame centered on the Si atom, representing the deformation modes of an isolated tetrahedron~\cite{taraskin97,oligschleger99,mukhopadhyay2003}---center in orange, vertices in red. Four modes capture bond-length changes: the non-degenerate $A_1$ mode corresponding to isotropic compression, and the triply degenerate $F_2^{\mathrm{s}}$ modes, corresponding to antagonistic bond-length changes, in which a pair of vertices stretches while the other pair contracts. Five modes capture bond-angle changes (bending): the triply degenerate $F_2^{\mathrm{b}}$ modes involve opposing angular changes of two pairs of vertices, while the doubly degenerate $E$ modes correspond to symmetric bending and torsion [see later the discussion of Eq.~\eqref{eq:E} for details].}
\end{figure}
Of particular interest here is the decomposition based on tetrahedral units, which includes both rigid-unit rotations (the $F_1$ modes) and internal shape degrees of freedom. The shape of a tetrahedron is parametrized by nine degrees of freedom, which organize into four irreducible symmetry-invariant subspaces~\cite{decius1977,taraskin97,shcheblanov16}, as illustrated in Fig.~\ref{fig:tetra}: a non-degenerate subspace ($A_1$) associated with isotropic (or symmetric) bond stretching; a triply degenerate subspace ($F_2^\mathrm{s}$) corresponding to antagonistic (or antisymmetric) stretching of vertex pairs; another triply degenerate subspace ($F_2^\mathrm{b}$) associated with antagonistic bond-angle changes (bending) between opposite vertex pairs; and finally, a doubly degenerate subspace ($E$) capturing symmetric bending and torsion. Since the symmetry-adapted modes are orthogonal to rigid translations of the considered unit, they are usually displayed in a frame centered on the Si atom, as in Fig.~\ref{fig:tetra}. Complete representations of the relevant local modes in polarization space, including the silicon component, will be given later when they arise naturally in the discussion.

The results of the local projection analysis are shown in Fig.~\ref{fig:local}-(b). As in the decomposition of atomic contributions in panel~(a), some trends are apparent: the low-frequency range is largely dominated by rigid-unit rotations, while the high-frequency doublet involves a prominent antisymmetric stretch contribution.

However, it is essential to note that this procedure does not decompose eigenvectors into orthogonal subspaces of the full polarization vector space $\mathbf{E}$ and therefore cannot lead to the identification of eigenspaces of the complete vibrational problem. This limitation is evident from a simple dimensional argument: in total, $4N$ local degrees of freedom are associated with tetrahedral units---$N$ for rotations and $3N$ for deformations---whereas the polarization space $\mathbf{E}$ has dimension $3N$. This inconsistency arises because the motions of oxygen atoms are counted twice, as each oxygen atom belongs to two tetrahedra. As a result, this approach provides only qualitative indications of how local symmetry-adapted modes contribute to vibrational eigenvectors at a given frequency. Like the decomposition based on atomic degrees of freedom, it does not isolate distinct families of atomic displacements that can be unambiguously associated with specific spectral features.

\section{Hierarchical analysis of silica vibrational modes}
\label{sec:decomposition}
In the following, we introduce several subspaces of $\mathbf{E}$, denoted by symbols of the form $\mathbf{E}_\mathrm{a}^\mathrm{b}$ with various subscripts, superscripts, and possibly accents. Systematically, the orthogonal projector onto $\mathbf{E}_\mathrm{a}^\mathrm{b}$, the restricted VDOS on $\mathbf{E}_\mathrm{a}^\mathrm{b}$, and the corresponding partial VDOS carry the same indices and accents, yet with their reserved symbols, i.e., are denoted $\mathbfcal{P}_\mathrm{a}^\mathrm{b}$, $\rho_\mathrm{a}^\mathrm{b}$, and $\varrho_\mathrm{a}^\mathrm{b}$, respectively.

\begin{table}[t]
\centering
\begin{tabular}{lcll}
\toprule
{Subspace} & {Dimension} & {Projector} & {Projector expression}\\
\hhline{====}
  $\mathbf{E}_\mathrm{bs}$ &$4N/3$& $\mathbfcal{P}_\mathrm{bs}$ & $\mathbfcal{A}\mathbfcal{A}^+$\\
  $\mathbf{E}_\mathrm{ns}$ &$5N/3$& $\mathbfcal{P}_\mathrm{ns}$ & $\mathbfcal{I}-\mathbfcal{P}_\mathrm{bs}=\mathbfcal{A}_\mathrm{ns}\mathbfcal{A}_\mathrm{ns}^+$\\
  $\mathbf{E}_{F_2^\mathrm{b}}$ &$N$& $\mathbfcal{P}_{F_2^\mathrm{b}}$ & $\mathbfcal{A}_{F_2^\mathrm{b}}\mathbfcal{A}_{F_2^\mathrm{b}}^+$\\
  $\mathbf{E}_{E}$ &$2N/3$& $\mathbfcal{P}_{E}$ & $\mathbfcal{A}_{E}\mathbfcal{A}_{E}^+$\\
\midrule
  $\mathbf{E}_\mathrm{ss}$ &$2N/3$& $\mathbfcal{P}_\mathrm{ss}$ & $\mathbfcal{A}_\mathrm{ss}\mathbfcal{A}_\mathrm{ss}^+$\\
  $\mathbf{E}_\mathrm{as}$ &$2N/3$& $\mathbfcal{P}_\mathrm{as}$ & $\mathbfcal{A}_\mathrm{as}\mathbfcal{A}_\mathrm{as}^+$\\
  $\mathbf{E}_\mathrm{as}^\perp$ &$2N/3$& $\mathbfcal{P}_\mathrm{as}^\perp$ & $\mathbfcal{P}_\mathrm{bs}\left(\mathbfcal{I}-\mathbfcal{P}_\mathrm{as}\right)$ \\
  $\mathbf{E}_\mathrm{dev}$ &$N$& $\mathbfcal{P}_\mathrm{dev}$ & $\mathbfcal{A}_\mathrm{dev}\mathbfcal{A}_\mathrm{dev}^+$ \\
  $\mathbf{E}_\mathrm{dev}^\perp$ &$N/3$&& $\mathbfcal{P}_\mathrm{bs}\left(\mathbfcal{I}-\mathbfcal{P}_\mathrm{dev}\right)$\\
\midrule
  $\widetilde{\mathbf{E}}_\mathrm{as}$ &$N/3$& $\widetilde{\mathbfcal{P}}_\mathrm{as}$& $\mathbfcal{P}_\mathrm{as}\wedge\mathbfcal{P}_\mathrm{dev}$\\
  $\widetilde{\mathbf{E}}_\mathrm{as}^\perp$ &$N/3$&& $\mathbfcal{P}_\mathrm{as}(\mathbfcal{I}-\widetilde{\mathbfcal{P}}_\mathrm{as})$\\
  $\widetilde{\mathbf{E}_\mathrm{as}^\perp}$ &$N/3$& $\widetilde{\mathbfcal{P}_\mathrm{as}^\perp}$& $(\mathbfcal{I}-\mathbfcal{P}_\mathrm{as})\wedge\mathbfcal{P}_\mathrm{dev}$\\
  $\widetilde{\mathbf{E}_\mathrm{as}^\perp}^\perp$ &$N/3$&& $\mathbfcal{P}_\mathrm{bs}(\mathbfcal{I}-\mathbfcal{P}_\mathrm{as})(\mathbfcal{I}-\widetilde{\mathbfcal{P}_\mathrm{as}^\perp})$\\
  $\widetilde{\mathbf{E}}_\mathrm{ss}$ &$N/3$& $\widetilde{\mathbfcal{P}}_\mathrm{ss}$ & $\mathbfcal{P}_\mathrm{ss}\wedge\mathbfcal{P}_\mathrm{dev}$\\
%  $\widetilde{\mathbf{E}}_\mathrm{ss}^\perp$ &$N/3$& $\widetilde{\mathbfcal{P}}_\mathrm{ss}^\perp$ & $\mathbfcal{P}_\mathrm{ss}\wedge(\mathbfcal{I}-\widetilde{\mathbfcal{P}}_\mathrm{ss})$\\
\midrule
  $\mathbf{E}_{\mathrm{ns},E}$ &$2N/3$&$\mathbfcal{P}_{\mathrm{ns},E}$& $\mathbfcal{P}_\mathrm{ns}\wedge(\mathbfcal{I}-\mathbfcal{P}_{F_2^\mathrm{b}})$\\
  $\mathbf{E}_{\mathrm{ns},E}^\perp$ &$N$&& $\mathbfcal{P}_\mathrm{ns}\,(\mathbfcal{I}-\mathbfcal{P}_{\mathrm{ns},E})$\\
  $\mathbf{E}_{\mathrm{ns},F_2^\mathrm{b}}$ &$N$&$\mathbfcal{P}_{\mathrm{ns},F_2^\mathrm{b}}$& $\mathbfcal{P}_\mathrm{ns}\wedge(\mathbfcal{I}-\mathbfcal{P}_{E})$\\
  $\mathbf{E}_{\mathrm{ns},F_2^\mathrm{b}}^\perp$ &$2N/3$&& $\mathbfcal{P}_\mathrm{ns}\,(\mathbfcal{I}-\mathbfcal{P}_{\mathrm{ns},F_2^\mathrm{b}})$\\
\bottomrule
\end{tabular}
\caption{\label{table:projectors} List of key subspaces, their dimensions, associated projection operator symbols (when defined), and reduced projector expressions (see text for details). The table is organized into four groups according to the parent space and splitting level. The first group contains projectors that split the full polarization space. The second and third groups correspond to second and third level splittings of the bond-stretch subspace, respectively. The fourth group corresponds to the splitting of the no-stretch subspace.}
\end{table}

Since the projectors can be constructed in multiple ways---depending on whether the associated subspace is defined as the image or co-kernel of a given matrix, or as the intersection of subspaces, in which case the expression depends on whether the corresponding projectors commute (see Appendix~\ref{app:projectors})---we provide for reference, in Table~\ref{table:projectors}, a complete list of the subspaces encountered in the decomposition of vitreous silica, together with the reduced expressions of the associated orthogonal projectors.

\subsection{First splitting: \ce{Si-O} bond stretches}
\label{sec:nsbs}
Let us start by using the projection formalism to isolate, within the vibrational problem, the most rigid elements of the microstructure---the \ce{Si-O} bonds. Displacements preserving the lengths of the \ce{Si-O} bonds to first order satisfy the constraint $\vec r_{ij}\cdot\vec u_{ij}=0$, for all bonded \ce{Si-O} pairs $(i,j)$, where $\vec r_{ij}=\vec r_j-\vec r_i$ and $\vec u_{ij}=\vec u_j-\vec u_i$. Each of these $4N/3$ constraints can be written in the form $\mathbf{L}^{(ij)}\cdot\mathbf{u}=0$, where $\mathbf{L}^{(ij)}=\mathbf{L}^{(ji)}$ is the displacement field whose only nonzero components are $\vec L^{(ij)}_i=-\vec r_{ij}$ and $\vec L^{(ij)}_j=\vec r_{ij}$. Introducing the $(3N)\times(4N/3)$ matrix $\mathbfcal{L}$ whose columns are the vectors $\mathbf{L}^{(ij)}$, the condition that no \ce{Si-O} bond changes length to first order becomes $\mathbfcal{L}^\intercal\,\mathbf{u}=\mathbf{0}$. Expressed in terms of the polarization vector $\mathbf{e}=\mathbfcal{M}^{1/2}\,\mathbf{u}$ this condition reads $\mathbfcal{A}^\intercal\,\mathbf{e}=\mathbf{0}$ with $\mathbfcal{A}=\mathbfcal{M}^{-1/2}\,\mathbfcal{L}$.

We denote by $\mathbf{E}_\mathrm{ns}=\mathrm{Ker}(\mathbfcal{A}^\intercal)$ the "no-stretch" (or "n.s.") subspace of polarization fields satisfying these constraints. Its orthogonal complement, the "bond-stretch" (or "b.s.") subspace, is therefore the image space of $\mathbfcal{A}$, namely $\mathbf{E}_\mathrm{bs}=\mathrm{Im}(\mathbfcal{A})$. The vectors $\mathbf{L}^{(ij)}$ are linearly independent since only two of them contribute to any oxygen displacement, with these contributions being generally non-parallel. It follows that $\mathbfcal{A}$ has rank $4N/3$, and therefore $\dim (\mathbf{E}_\mathrm{bs})=4N/3$ and $\dim (\mathbf{E}_\mathrm{ns})=5N/3$.

\begin{figure}[t]
  \medskip
  \includegraphics[width=0.47\textwidth]{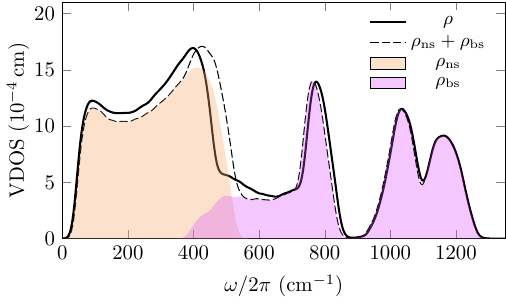}
  \caption{\label{fig:first} VDOS of the full system ($\rho$, black solid) compared with those of the n.s.\ ($\rho_\mathrm{ns}$, orange area) and b.s.\ ($\rho_\mathrm{bs}$, purple area) subspaces. The spectrum of the uncoupled system ($\rho_\mathrm{ns}+\rho_\mathrm{bs}$, black dashed) remains close to the full spectrum $\rho$, except within the narrow frequency range where $\rho_\mathrm{ns}$ and $\rho_\mathrm{bs}$ overlap. This demonstrates that the full VDOS results from a moderate and essentially resonant hybridization between the n.s.\ and b.s.\ eigenmodes.}
\end{figure}

To assess the relevance of this decomposition, we numerically diagonalize the vibrational problems restricted to the n.s.\ and b.s.\ subspaces after constructing the corresponding projectors, $\mathbfcal{P}_\mathrm{bs}=\mathbfcal{A}\mathbfcal{A}^+$ and $\mathbfcal{P}_\mathrm{ns}=\mathbfcal{I}-\mathbfcal{P}_\mathrm{bs}$. The corresponding spectra, $\rho_\mathrm{ns}$ and $\rho_\mathrm{bs}$, are shown in Fig.~\ref{fig:first} along with the full spectrum. Although they overlap over a narrow frequency interval, the two restricted spectra clearly occupy largely distinct domains: the n.s.\ subspace captures the entire low-frequency region, while the b.s.\ subspace accounts for the high-frequency part of the spectrum, above the overlap region.

This separation justifies viewing the full spectrum as resulting from the recoupling of the n.s.\ and b.s.\ vibrational problems through the off-diagonal blocks of $\mathbfcal{D}$ in Eq.~\eqref{eq:decomposition}. Notably, the spectrum of the uncoupled system, $\rho_\mathrm{ns}+\rho_\mathrm{bs}$, differs from the full spectrum $\rho$ primarily in the overlap region, indicating that the hybridization between n.s.\ and b.s.\ modes is weak and essentially resonant. Outside this narrow resonant band, recoupling induces only slight downward and upward shifts of the peaks near \SI{400}{cm^{-1}} and \SI{800}{cm^{-1}}, respectively, reflecting minimal level repulsion.

\subsection{The bond-stretch subspace}
\label{sec:bs}
By definition, any element $\mathbf{e}\in\mathbf{E}_\mathrm{bs}$ of the b.s.\ subspace can be written as $\mathbf{e}=\mathbfcal{A}\,\boldsymbol{\alpha}$, where $\boldsymbol{\alpha}$ is a coefficient vector determining elementary bond-stretch contributions. The corresponding displacement field $\mathbf{u}=\mathbfcal{M}^{-1/2}\,\mathbf{e}=\mathbfcal{M}^{-1}\mathbfcal{L}\boldsymbol{\alpha}$ is therefore a linear combination of the vectors $\mathbfcal{M}^{-1}\mathbf{L}^{(ij)}$. Any such displacement field preserves the center of mass. This applies, in particular, when $\boldsymbol{\alpha}$ has only a few non-zero components, and hence for any elementary bond-stretch field generated by a structural subunit composed of \ce{Si-O} bonds---which includes \ce{Si-O} bonds, \ce{Si-O-Si} bridges, and tetrahedra. This property results from $\mathbf{L}^{(ij)}$ being orthogonal to rigid translations, since it is the configuration-space gradient of a translation-invariant quantity, $r_{ij}^2/2$.\\

The spectrum of $\mathbf{E}_\mathrm{bs}$ exhibits clearly separated bands, with a large gap around \SI{900}{cm^{-1}}. Such a separation suggests the existence of a pronounced stiffness contrast between different classes of atomic displacements. To identify the corresponding subspaces, we exploit the fact that $\mathbf{E}_\mathrm{bs}$ is the image of $\mathbfcal{A}$, so that its elements can be expressed explicitly as linear combinations of column vectors. We denote by $\mathbf{A}^{(ij)}$ the column vectors of $\mathbfcal{A}$, where $(i,j)$ runs over bonded \ce{Si-O} pairs.

\subsubsection{Secondary splitting: symmetric and antisymmetric stretches}\label{sec:asss}
Let us consider a \ce{Si-O-Si} triplet with atoms labeled $i,j,k$, respectively. In the b.s.\ subspace, the polarization field at the central oxygen atom---and therefore its displacement---arises solely from the combination of the vectors $A_j^{(ij)}=\vec r_{ij}/\sqrt{m_j}$ and $A_j^{(kj)}=\vec r_{kj}/\sqrt{m_j}$. Consequently, this displacement lies in the \ce{Si-O-Si} plane and has no rocking component. Since the \ce{Si-O-Si} angles systematically lie in the range $[\pi/2,\pi]$ (see Fig.~\ref{fig:model}-(a) and Refs.~\cite{giacomazzi09,trease2017}), there is a systematic stiffness contrast between displacements in which the oxygen atom moves transverse to the \ce{Si-Si} axis (softer) or parallel to it (stiffer)~\cite{bock70,sen77,galeener79,bks1990}.

\begin{figure}[t]
  \begin{center}
    \includegraphics[width=0.12\textwidth]{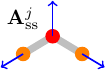}\hfil
    \includegraphics[width=0.13\textwidth]{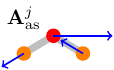}
  \end{center}
  \caption{\label{fig:asssmodes}
    Non-zero vector component of generating polarization vectors of symmetric and antisymmetric stretches, $\mathbf{A}_\mathrm{ss}^j$ and  $\mathbf{A}_\mathrm{as}^j$ (up to an arbitrary global scale), for an \ce{Si-O-Si} triplet---silicon atoms in orange and oxygen in red.}
\end{figure}

This stiffness contrast motivates decomposing the $\mathbf{E}_\mathrm{bs}$ spectrum by distinguishing transverse and parallel oxygen displacements relative to the \ce{Si-Si} axis. This construction is analogous to the decomposition of oxygen displacements into bending and stretching components illustrated in Fig.~\ref{fig:local}, but must be performed within the polarization subspace $\mathbf{E}_\mathrm{bs}$. To this end, for any \ce{Si-O-Si} triplet with respective indices $(i,j,k)$, we define the \emph{symmetric stretch} (s.s.) field,
\begin{equation}
  \mathbf{A}_\mathrm{ss}^j=\mathbf{A}^{(ij)}+\mathbf{A}^{(kj)}\qquad,
\end{equation}
and the \emph{antisymmetric stretch} (a.s.) field,
\begin{equation}
  \mathbf{A}_\mathrm{as}^j=\mathbf{A}^{(ij)}-\mathbf{A}^{(kj)}\qquad.
\end{equation}
The non-zero components of these $3N$-vectors are supported only by the atoms of the \ce{Si-O-Si} triplet and are schematically represented in Fig.~\ref{fig:asssmodes}.

Symmetric and antisymmetric stretch fields generate subspaces denoted $\mathbf{E}_\mathrm{ss}=\mathrm{Im}(\mathbfcal{A}_\mathrm{ss})$ and $\mathbf{E}_\mathrm{as}=\mathrm{Im}(\mathbfcal{A}_\mathrm{as})$, with $\mathbfcal{A}_\mathrm{ss}$ and $\mathbfcal{A}_\mathrm{as}$ being the $(3N)\times(2N/3)$ matrices whose columns are $\{\mathbf{A}_\mathrm{ss}^j\}$ and $\{\mathbf{A}_\mathrm{as}^j\}$, respectively. In $\mathbf{E}_\mathrm{as}$, since $A_{\mathrm{as}\,j}^{j}=\vec r_{ik}/\sqrt{m_j}$, oxygen $j$ moves along the corresponding \ce{Si-Si} axis, which is the stiffest direction for its motion. In contrast, in $\mathbf{E}_\mathrm{ss}$, it tends to move transversely to the \ce{Si-Si} axis---this property holds approximately in general and becomes exact only when the two adjacent \ce{Si-O} bond lengths are equal, which is a good approximation for silica where bond lengths are narrowly distributed (see Fig.~\ref{fig:model}-(b) and Refs.~\cite{giacomazzi09,trease2017,srivastava2018}). Additionally, all fields in set $\{\mathbf{A}_\mathrm{ss}^j\}$ (resp. $\{\mathbf{A}_\mathrm{as}^j\}$) are mutually linearly independent, since each is supported on a distinct oxygen atom. Consequently, both matrices $\mathbfcal{A}_\mathrm{ss}$ and $\mathbfcal{A}_\mathrm{as}$ are full-column-rank and $\dim(\mathbf{E}_\mathrm{ss})=\dim(\mathbf{E}_\mathrm{as})=2N/3$.

This splitting differs in two key respects from the separation between the bending and stretching components of oxygen displacements [Fig.~\ref{fig:local}-(a)]~\cite{dean72,galeener83,taraskin97}:
\begin{compactitem}
\item it does not isolate oxygen motion, since it is defined within the b.s.\ subspace, whose generating vectors, the fields $\mathbf{A}^{(ij)}$, combine antagonistic displacements of both \ce{Si} and \ce{O} with weights determined by the n.s./b.s.\ splitting;
\item consequently, as illustrated in Fig.~\ref{fig:asssmodes}, the generating vectors $\mathbf{A}_\mathrm{ss}^j$ and $\mathbf{A}_\mathrm{as}^j$ involve coupled motion of all three atoms in each \ce{Si-O-Si} triplet.
\end{compactitem}
It follows that $\mathbf{E}_\mathrm{ss}$ and $\mathbf{E}_\mathrm{as}$ are \emph{not} mutually orthogonal. Indeed, for two oxygen atoms $j\ne j'$ bonded to a common silicon atom $i$, one finds $\mathbf{A}_\mathrm{ss}^j\cdot\mathbf{A}_\mathrm{as}^{j'}=\pm\vec r_{ij}\cdot\vec r_{ij'}/m_i\ne0$. Therefore, the b.s.\ spectrum cannot be decomposed orthogonally into $\mathbf{E}_\mathrm{ss}$ and $\mathbf{E}_\mathrm{as}$. Instead, two alternative decompositions are available: $\mathbf{E}_\mathrm{bs}=\mathbf{E}_\mathrm{ss}\oplus\mathbf{E}_\mathrm{ss}^\perp$ or $\mathbf{E}_\mathrm{bs}=\mathbf{E}_\mathrm{as}\oplus\mathbf{E}_\mathrm{as}^\perp$.

As with the initial splitting $\mathbf{E}=\mathbf{E}_\mathrm{ns}\oplus \mathbf{E}_\mathrm{bs}$ based on bond-stretch constraints, we expect to be able to identify relevant subspaces by isolating the stiffest contributions to the considered vibrational problem. Within $\mathbf{E}_\mathrm{bs}$, these correspond to antisymmetric stretches. Accordingly, we tentatively examine the orthogonal decomposition $\mathbf{E}_\mathrm{bs}=\mathbf{E}_\mathrm{as}\oplus \mathbf{E}_\mathrm{as}^\perp$.

For convenience, we do not attempt to explicitly construct vectors and operators intrinsic to the subspace $\mathbf{E}_\mathrm{bs}$. Instead, we work with polarization vectors and operators defined on the full space $\mathbf{E}$ and use the projection formalism to isolate the vibrational problem restricted to $\mathbf{E}_\mathrm{bs}$. Since $\mathbf{E}_\mathrm{as}$ is generated by linear combinations of the vectors $\mathbf{A}^{(ij)}$, it follows that $\mathrm{Im}(\mathbfcal{A}_\mathrm{as})=\mathbf{E}_\mathrm{as}\subset\mathbf{E}_\mathrm{bs}$. Consequently, the orthogonal projection onto $\mathbf{E}_\mathrm{as}$ is given by $\mathbfcal{P}_\mathrm{as}=\mathbfcal{A}_\mathrm{as}\mathbfcal{A}_\mathrm{as}^+$ and satisfies $\mathbfcal{P}_\mathrm{as}\mathbfcal{P}_\mathrm{bs}=\mathbfcal{P}_\mathrm{as}=\mathbfcal{P}_\mathrm{bs}\mathbfcal{P}_\mathrm{as}$. By contrast, the operator $\mathbfcal{I}-\mathbfcal{P}_\mathrm{as}$ projects onto the orthogonal complement of $\mathbf{E}_\mathrm{as}$ in the full space $\mathbf{E}$, not onto its complement within $\mathbf{E}_\mathrm{bs}$. The orthogonal projector onto the complementary subspace $\mathbf{E}_\mathrm{as}^\perp\subset\mathbf{E}_\mathrm{bs}$ is therefore $(\mathbfcal{I}-\mathbfcal{P}_\mathrm{as})\wedge\mathbfcal{P}_\mathrm{bs}=(\mathbfcal{I}-\mathbfcal{P}_\mathrm{as})\mathbfcal{P}_\mathrm{bs}=\mathbfcal{P}_\mathrm{bs}(\mathbfcal{I}-\mathbfcal{P}_\mathrm{as})$, where the wedge operator simplifies to the product because the involved projectors commute [see Appendix~\ref{app:projectors}].

\begin{figure}[t]
  \includegraphics[width=0.47\textwidth]{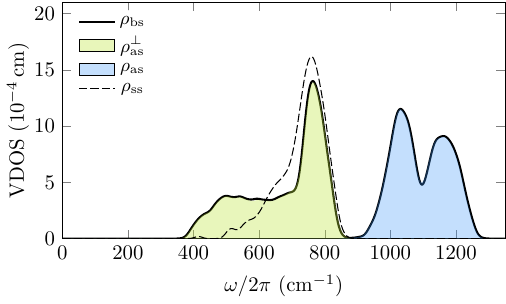}
  \caption{\label{fig:ps:second}
    Orthogonal decomposition of the b.s.\ subspace $\mathbf{E}_\mathrm{bs}=\mathbf{E}_\mathrm{as}\oplus\mathbf{E}_\mathrm{as}^\perp$, with $\rho_\mathrm{as}$ and $\rho_\mathrm{as}^\perp$ denoting the spectra of the uncoupled subspaces $\mathbf{E}_\mathrm{as}$ and $\mathbf{E}_\mathrm{as}^\perp$, respectively. Also shown is the spectrum of the uncoupled s.s.\ subspace, $\rho_\mathrm{ss}$ (dashed line), which lies at the center of the b.s.\ spectrum.}
\end{figure}

The spectra of $\mathbf{E}_\mathrm{as}$ and $\mathbf{E}_\mathrm{as}^\perp$, shown in Fig.~\ref{fig:ps:second}, nearly perfectly match the two bands of the b.s.\ spectrum. A very small overlap remains near the gap, but it is too small to be distinguishable on this plot and may arise from finite-size effects. This demonstrates that $\mathbf{E}_\mathrm{as}$ and $\mathbf{E}_\mathrm{as}^\perp$ correspond almost perfectly to the eigenspaces of $\mathbfcal{D}_\mathrm{bs}$ associated with the high-frequency doublet and the mid-frequency band, respectively. As a corollary, each band contains $2N/3$ modes. Since $\mathbf{E}_\mathrm{as}=\mathrm{Im}(\mathbfcal{A}_\mathrm{as})$ is generated by the anti-symmetric stretch vector fields $\{\mathbf{A}_\mathrm{as}^j\}$, we have also obtained an explicit expression for the modes forming the high-frequency doublet of silica, which is a significant result in itself.

Note that although $\mathbf{E}_\mathrm{as}^\perp\ne\mathbf{E}_\mathrm{ss}$ since $\mathbf{E}_\mathrm{ss}$ and $\mathbf{E}_\mathrm{as}$ are not mutually orthogonal, both $\mathbf{E}_\mathrm{as}^\perp$ and $\mathbf{E}_\mathrm{ss}$ have the same dimension, $2N/3$. Moreover, linear combinations of vectors from $\mathbf{E}_\mathrm{ss}$ and $\mathbf{E}_\mathrm{as}$ generate the entire b.s.\ subspace. Therefore, one may view $\mathbf{E}_\mathrm{as}^\perp$ as being obtained by taking vectors from $\mathbf{E}_\mathrm{ss}$ and projecting out their a.s.\ component. From this perspective, the stiffness contrast between $\mathbf{E}_\mathrm{as}$ and $\mathbf{E}_\mathrm{as}^\perp$ should reflect in part the contrast between the a.s.\ and s.s.\ polarization vectors. This interpretation is supported by the dashed curve in Fig.~\ref{fig:ps:second}, which shows the intrinsic spectrum of the uncoupled s.s.\ problem: as expected, it differs from that of $\mathbf{E}_\mathrm{as}^\perp$, but lies entirely within the same frequency range and peaks around \SI{800}{cm^{-1}}, indicating that the s.s.\ modes contribute significantly to the main peak of the mid-frequency band.

A striking feature of the mid-frequency band is the presence of a shoulder extending to lower frequencies than the s.s.\ spectrum. This feature therefore must arise from orthogonality constraints defining $\mathbf{E}_\mathrm{as}^\perp$: compared with $\mathbf{E}_\mathrm{ss}$ modes, these constraints eliminate the stiff a.s.\ component, thereby shifting the frequency content of the mid-band compared with the s.s.\ spectrum. To understand qualitatively how orthogonality constraints generate lower-frequency modes than those of the uncoupled s.s.\ problem, consider an arbitrary eigenmode $\mathbf{e}_\mathrm{as}^\perp\in\mathbf{E}_\mathrm{as}^\perp$ of the restricted b.s.\ problem governed by $\mathbfcal{D}_\mathrm{bs}=\mathbfcal{P}_\mathrm{bs}\mathbfcal{D}\mathbfcal{P}_\mathrm{bs}$ with eigenfrequency $\omega$, and decompose it as $\mathbf{e}_\mathrm{as}^\perp=\mathbf{e}_\mathrm{ss}+\mathbf{e}_\mathrm{as}$ with $\mathbf{e}_\mathrm{ss}\in\mathbf{E}_\mathrm{ss}$ and $\mathbf{e}_\mathrm{as}\in\mathbf{E}_\mathrm{as}$. Assuming the mode normalized, $\|\mathbf{e}_\mathrm{as}^\perp\|{^2}=1$, and using its orthogonality to $\mathbf{e}_\mathrm{as}$, we find $\|\mathbf{e}_\mathrm{ss}\|^2=1+\|\mathbf{e}_\mathrm{as}\|{^2}$. Using the fact that $\mathbf{E}_\mathrm{as}$ and $\mathbf{E}_\mathrm{as}^\perp$ are eigenspaces of $\mathbfcal{D}_\mathrm{bs}$, we obtain:
\begin{equation}
\mathbf{e}_\mathrm{ss}^\intercal\mathbfcal{D}_\mathrm{bs}\,\mathbf{e}_\mathrm{ss}=\omega^2+\mathbf{e}_\mathrm{as}^\intercal\mathbfcal{D}_\mathrm{bs}\,\mathbf{e}_\mathrm{as}\,,
\end{equation}
and after introducing the Rayleigh quotients,
\begin{equation}
  \begin{split}
    \omega_\mathrm{ss}^2&=\frac{\mathbf{e}_\mathrm{ss}^\intercal\mathbfcal{D}_\mathrm{bs}\,\mathbf{e}_\mathrm{ss}}{\|\mathbf{e}_\mathrm{ss}\|^2} \,,\\
    \omega_\mathrm{as}^2&=\frac{\mathbf{e}_\mathrm{as}^\intercal\mathbfcal{D}_\mathrm{bs}\,\mathbf{e}_\mathrm{as}}{\|\mathbf{e}_\mathrm{as}\|^2} \,,\\
  \end{split}
\end{equation}
we find:
\begin{equation}\label{eq:omega}
\begin{split}
\omega^2&=\omega_\mathrm{ss}^2\|\mathbf{e}_\mathrm{ss}\|^2-\omega_\mathrm{as}^2\|\mathbf{e}_\mathrm{as}\|{^2}\\
&=\omega_\mathrm{ss}^2+(\omega_\mathrm{ss}^2-\omega_\mathrm{as}^2)\|\mathbf{e}_\mathrm{as}\|{^2} \, .
\end{split}
\end{equation}
Note that $\mathbf{e}_\mathrm{ss}$ is not an eigenmode of $\mathbfcal{D}_\mathrm{bs}$ since $\mathbf{E}_\mathrm{ss}$ is not an eigenspace of this operator. However, since $\mathbf{e}_\mathrm{ss}\in\mathbf{E}_\mathrm{ss}$, we have $\mathbf{e}_\mathrm{ss}^\intercal\mathbfcal{D}_\mathrm{bs}\,\mathbf{e}_\mathrm{ss}=\mathbf{e}_\mathrm{ss}^\intercal\mathbfcal{P}_\mathrm{ss}\mathbfcal{D}_\mathrm{bs}\mathbfcal{P}_\mathrm{ss}\,\mathbf{e}_\mathrm{ss}$, where $\mathbfcal{P}_\mathrm{ss}\mathbfcal{D}_\mathrm{bs}\mathbfcal{P}_\mathrm{ss}$ is the restriction of $\mathbfcal{D}_\mathrm{bs}$ on $\mathbf{E}_\mathrm{ss}$. The Rayleigh quotient $\omega_\mathrm{ss}^2$ is therefore a weighted average of the eigenvalues of $\mathbfcal{P}_\mathrm{ss}\mathbfcal{D}_\mathrm{bs}\mathbfcal{P}_\mathrm{ss}$, i.e., of the squared frequencies of the uncoupled s.s.\ subspace. The same holds for $\omega_\mathrm{as}^2$ since $\mathbf{E}_\mathrm{as}$ is an eigenspace of $\mathbfcal{D}_\mathrm{bs}$. It follows that $\omega_\mathrm{ss}^2<\omega_\mathrm{as}^2$, and from Eq.~\eqref{eq:omega}, that the eigenfrequency $\omega$ is shifted downward from $\omega_\mathrm{ss}$.

This explains how the orthogonal projection of the s.s.\ subspace away from the a.s.\ subspace, yielding $\mathbf{E}_\mathrm{as}^\perp$, produces systematic downward shifts of the intrinsic s.s.\ frequencies. These shifts arise not from hybridization but from orthogonality constraints, and give rise to the low-frequency shoulder in the mid-frequency band.

\subsubsection{Alternative secondary splitting: tetrahedral stretches}
Let us observe that $\rho_\mathrm{as}$ and $\rho_\mathrm{as}^\perp$ exhibit pronounced substructure, suggesting that further splitting is possible. Namely, the mid-frequency band ($\rho_\mathrm{as}^\perp$) comprises a sharp peak around \SI{800}{cm^{-1}} and a shoulder extending to lower frequencies, while the high-frequency band ($\rho_\mathrm{as}$) forms a doublet. Analysis of the third-level splitting responsible for these features requires introducing an alternative decomposition of the bond-stretch subspace $\mathbf{E}_\mathrm{bs}$ that separates \emph{isotropic} and \emph{deviatoric} stretch contributions to tetrahedral strains. The present subsection develops the formal construction of this decomposition, which is a necessary step before returning to the analysis of $\rho_\mathrm{as}$ and $\rho_\mathrm{as}^\perp$.

Consider an arbitrary b.s.\ polarization vector, $\mathbf{e}=\mathbfcal{A} \,\boldsymbol{\alpha}\in\mathbf{E}_\mathrm{bs}$, and group its components $\boldsymbol{\alpha}=\{\alpha^{(ij)}\}$ into quadruplets associated with each silicon atom: $\boldsymbol{\alpha}^{(i)}=\{{\alpha}_a^{(i)},a=1,\ldots,4\}=\{\alpha^{(ij)},j=j_1,\ldots,j_4\}$, where indices $j_a$ label the four oxygen atoms bonded to silicon atom $i$. Each quadruplet can be transformed into \emph{tetrahedral stretches} $\mathbf{T}^{(i)}=\{T_a^{(i)},a=1,\ldots,4\}=\mathbf{K}\,\boldsymbol{\alpha}^{(i)}$ using the matrix:
\begin{equation}
  \mathbf{K}=\frac{1}{2}
  \begin{pmatrix}
    1& 1& 1& 1\\
    1& 1& -1& -1\\
    1& -1& 1& -1\\
    1& -1& -1& 1\\
  \end{pmatrix} \,,
\end{equation}
which is symmetric and involutory (i.e., equal to its own inverse, since $\mathbf{K}^2=\mathbf{I}_4$, the $4\times4$ identity matrix). Thanks to the latter property, the original components can be recovered from the tetrahedral stretches as $\boldsymbol{\alpha}^{(i)}=\mathbf{K}\mathbf{T}^{(i)}$.

This local mapping between $\boldsymbol{\alpha}^{(i)}$ and $\mathbf{T}^{(i)}$ at tetrahedral scale extends naturally to the entire $\mathbf{E}_\mathrm{bs}$ subspace. To make it explicit, we introduce the generalized vector of tetrahedral stretches $\mathbf{T}=\{\mathbf{T}^{(i)}, i\in I_{\ce{Si}}\}$, where $I_{\ce{Si}}$ denotes the set of silicon atom indices, and we group the component vector $\boldsymbol{\alpha}$ into the corresponding quadruplets, $\boldsymbol{\alpha}=\{\boldsymbol{\alpha}^{(i)}, i\in I_{\ce{Si}}\}$. These vectors are related by $\boldsymbol{\alpha}=\mathbfcal{K}\mathbf{T}$ and $\mathbf{T}=\mathbfcal{K}\boldsymbol{\alpha}$, where $\mathbfcal{K}$ is the block-diagonal matrix composed of $N/3$ identical blocks equal to $\mathbf{K}$. The matrix $\mathbfcal{K}$ is symmetric and involutory ($\mathbfcal{K}^2=\mathbfcal{I}$).
%old{The matrix $\mathbfcal{K}$ is symmetric and orthogonal, satisfying $\mathbfcal{K}^2=\mathbfcal{I}$.}

At the scale of a single tetrahedron, we expect an elastic stiffness contrast between the \emph{isotropic} strain component $T_1^{(i)}$ and the three remaining components, which are deviatoric. The subspace $\mathbf{E}_\mathrm{iso}$ generated by the isotropic tetrahedral stretches has dimension $\dim(\mathbf{E}_\mathrm{iso})=N/3$, while the subspace $\mathbf{E}_\mathrm{dev}$ generated by deviatoric strains has dimension $\dim(\mathbf{E}_\mathrm{dev})=N$. Although these subspaces map orthogonally (via $\mathbfcal{K}$) onto the component vector $\boldsymbol{\alpha}$, they are not mutually orthogonal as subspaces of $\mathbf{E}_\mathrm{bs}$. To see this, it suffices to consider vectors from either subspace associated with two corner-sharing tetrahedra: their components on the shared oxygen generally yield a nonzero scalar product. This property follows from the fact that the column vectors of $\mathbfcal{A}$ are not mutually orthogonal.

The subspace of isotropic strains is too small to play a meaningful role in the following analysis. However, the space of the deviatoric strains is of central importance, and we explicitly construct its generating vectors. For each silicon atom $i$, these are given by
\begin{equation}
\begin{split}
  {\mathbf{A}}_{\mathrm{dev},1}^i&=\frac{1}{2}\left(\mathbf{A}^{(ij_1)}+\mathbf{A}^{(ij_2)}-\mathbf{A}^{(ij_3)}-\mathbf{A}^{(ij_4)}\right) \, ,\\
  {\mathbf{A}}_{\mathrm{dev},2}^i&=\frac{1}{2}\left(\mathbf{A}^{(ij_1)}-\mathbf{A}^{(ij_2)}+\mathbf{A}^{(ij_3)}-\mathbf{A}^{(ij_4)}\right) \, ,\\
  {\mathbf{A}}_{\mathrm{dev},3}^i&=\frac{1}{2}\left(\mathbf{A}^{(ij_1)}-\mathbf{A}^{(ij_2)}-\mathbf{A}^{(ij_3)}+\mathbf{A}^{(ij_4)}\right) \, ,
\end{split}
\end{equation}
where $j_1,\ldots,j_4$ denote the indices of the oxygen atoms bonded to silicon $i$. Let $\mathbfcal{A}_\mathrm{dev}$ denote the matrix composed of these $N$ column vectors. Then the deviatoric stretch subspace is given by $\mathbf{E}_\mathrm{dev}=\mathrm{Im}(\mathbfcal{A}_\mathrm{dev})$ and the corresponding orthogonal projector is $\mathbfcal{P}_\mathrm{dev}=\mathbfcal{A}_\mathrm{dev}\mathbfcal{A}_\mathrm{dev}^+$.\\

\begin{figure}[t]
  \includegraphics[width=0.47\textwidth]{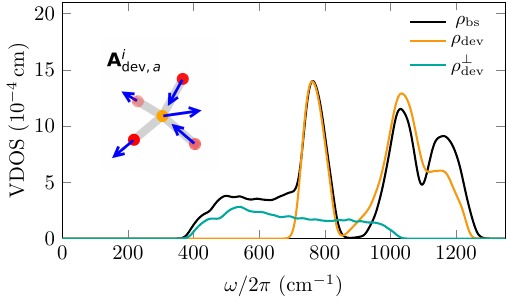}
  \caption{\label{fig:tetrahedral} Orthogonal decomposition of the b.s.\ subspace as $\mathbf{E}_\mathrm{bs}=\mathbf{E}_\mathrm{dev}\oplus\mathbf{E}_\mathrm{dev}^\perp$: $\rho_\mathrm{dev}$ and $\rho_\mathrm{dev}^\perp$ are the spectra of the uncoupled vibrational problems restricted to $\mathbf{E}_\mathrm{dev}$ and $\mathbf{E}_\mathrm{dev}^\perp$, respectively. Inset: sketch of an elementary deviatoric stretch mode, displaying all atomic contributions---silicon atom in orange, oxygen atoms in red.}
\end{figure}

The elementary tetrahedral stretch components $T_1^{(i)}$ and $\mathbf{T}_\mathrm{dev}^{(i)}=\{T_a^{(i)},a=2,\ldots,4\}$ defined above correspond to the symmetry-adapted deformation modes respectively labeled $A_1$ and $F_2^\mathrm{s}$~\cite{wilson96,pavlatou97,taraskin97,shcheblanov16,oligschleger99,mukhopadhyay2003} (or $T_2$~\cite{sarnthein97,pasquarello98a}) in the literature. However, the specific definition of these modes varies across studies. Some works define them relative to a fixed silicon atom~\cite{wilson96,pavlatou97,taraskin97,shcheblanov16,oligschleger99,mukhopadhyay2003}, while others require them to preserve the center of mass of the structural unit~\cite{sarnthein97,pasquarello98a}.

Here, as noted at the beginning of Sec.~\ref{sec:bs}, all bond-stretch polarization fields preserve the center of mass. Consequently, the polarization fields associated with tetrahedral stretch components also do, by construction. As a result, the deviatoric contributions $\mathbf{T}_\mathrm{dev}^{(i)}$, illustrated in the inset of Fig.~\ref{fig:tetrahedral}, differ significantly from the $F_2^{\mathrm{s}}$ modes, as defined in Refs.~\cite{taraskin97,shcheblanov16}, due to the coupled motion of oxygen and silicon atoms, with the latter carrying a significant non-zero contribution. This construction also slightly differs from approaches that enforce conservation of the center of mass for $T_2$ modes, but assume that the silicon atom motion is fixed in $A_1$~\cite{sarnthein97,pasquarello98a}, which is strictly correct only for a perfectly symmetric tetrahedron. By contrast, the present formulation is fully consistent with the bond-stretch constraints and naturally accounts for the silicon motion arising from deviations from perfect tetrahedral symmetry in all tetrahedral-stretch modes.\\

Figure~\ref{fig:tetrahedral} shows the spectra $\rho_\mathrm{dev}$ and $\rho_\mathrm{dev}^\perp$ of the uncoupled $\mathbf{E}_\mathrm{dev}$ and $\mathbf{E}_\mathrm{dev}^\perp$ subspaces, respectively. As expected, neither spectrum captures the bands of $\mathbf{E}_\mathrm{bs}$, since the primary stiffness contrast responsible for the separation between the mid-frequency band and the high-frequency doublet arises instead from antisymmetric stretches and their complements. Nevertheless, several notable features emerge: (i) the deviatoric contribution is unevenly distributed between the two peaks of the high-frequency doublet, being more pronounced in the lower-frequency peak, a feature previously observed by Sarnthein \etal~\cite{sarnthein97}; (ii) the peak around \SI{800}{cm^{-1}} originates predominantly from $\rho_\mathrm{dev}$; (iii) $\rho_\mathrm{dev}^\perp$ accounts for most of the low-frequency shoulder.

Therefore, although the stiffness contrast between tetrahedral stretches does not govern the primary band separation of the b.s.\ spectrum, it clearly contributes to its internal substructure.

\subsubsection{Third level splitting I: the high-frequency doublet}\label{sec:high}
In order to address this issue, we need to understand how the primary stiffness contrast captured in the decomposition $\mathbf{E}_\mathrm{bs}=\mathbf{E}_\mathrm{as}\oplus\mathbf{E}_\mathrm{as}^\perp$ combines with the secondary stiffness contrast captured by $\mathbf{E}_\mathrm{bs}=\mathbf{E}_\mathrm{dev}\oplus\mathbf{E}_\mathrm{dev}^\perp$.

In terms of the component vector $\boldsymbol{\alpha}$, the a.s.\ subspace is defined by the set of $2N/3$ constraints $\alpha^{(ij)}=-\alpha^{(kj)}$, for any oxygen $j$, where $i$ and $k$ denote the indices of the two silicon atoms it is bonded to. Since no non-zero element of $\mathbf{E}_\mathrm{iso}$ (dimension $N/3$) can satisfy this number of independent constraints, the intersection $\mathbf{E}_\mathrm{as}\cap\mathbf{E}_\mathrm{iso}$ reduces to the trivial (zero-dimensional) subspace $\{\mathbf{e}=\mathbf{0}\}$. By contrast, the subspace $\widetilde{\mathbf{E}}_\mathrm{as}\equiv\mathbf{E}_\mathrm{as}\cap\mathbf{E}_\mathrm{dev}$ has dimension $N-2N/3=N/3$ (since $N$ is the dimensions of $\mathbf{E}_\mathrm{dev}$ and $2N/3$ the number of constraints defining $\mathbf{E}_\mathrm{as}$), which is exactly half the dimension of ${\mathbf{E}}_\mathrm{as}$.
Therefore, the a.s.\ subspace can be decomposed into two $N/3$-dimensional complementary subspaces as $\mathbf{E}_\mathrm{as}=\widetilde{\mathbf{E}}_\mathrm{as}\oplus\widetilde{\mathbf{E}}_\mathrm{as}^\perp$. The tilde here and in the rest of the paper is used to denote the intersection of a subspace with $\mathbf{E}_\mathrm{dev}$.

\begin{figure}[t]
  \includegraphics[width=0.47\textwidth]{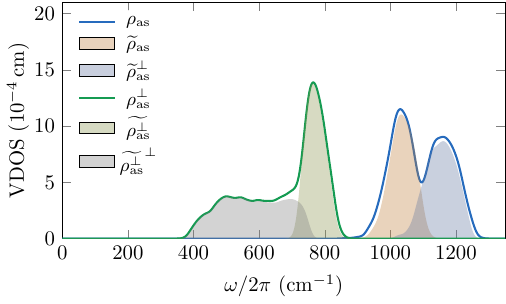}
  \caption{\label{fig:ps:third} Third level splitting for the b.s.\ subspace. The decomposition of the a.s.\ subspace as $\mathbf{E}_\mathrm{as}=\widetilde{\mathbf{E}}_\mathrm{as}\oplus\widetilde{\mathbf{E}}_\mathrm{as}^\perp$ is examined by comparing the a.s.\ spectrum $\rho_\mathrm{as}$ with the spectra $\widetilde\rho_\mathrm{as}$ and $\widetilde\rho_\mathrm{as}^\perp$ of the uncoupled subspaces $\widetilde{\mathbf{E}}_\mathrm{as}$ and $\widetilde{\mathbf{E}}_\mathrm{as}^\perp$, respectively. The decomposition $\mathbf{E}_\mathrm{as}^\perp=\widetilde{\mathbf{E}_\mathrm{as}^\perp}\oplus\widetilde{\mathbf{E}_\mathrm{as}^\perp}^\perp$ is examined analogously by comparing $\rho_\mathrm{as}^\perp$ with the spectra $\widetilde{\rho_\mathrm{as}^\perp}$ and $\widetilde{\rho_\mathrm{as}^\perp}^\perp$ of the uncoupled subspaces $\widetilde{\mathbf{E}_\mathrm{as}^\perp}$ and $\widetilde{\mathbf{E}_\mathrm{as}^\perp}^\perp$, respectively.}
\end{figure}

The orthogonal projector onto $\widetilde{\mathbf{E}}_\mathrm{as}$ is $\widetilde{\mathbfcal{P}}_\mathrm{as}=\mathbfcal{P}_\mathrm{as}\wedge\mathbfcal{P}_\mathrm{dev}$, with the $\wedge$ operator introduced in Sec.~\ref{sec:projectors} and explicitly constructed in Appendix~\ref{app:projectors}. The orthogonal projector onto $\widetilde{\mathbf{E}}_\mathrm{as}^\perp$ is $\widetilde{\mathbfcal{P}}_\mathrm{as}^\perp=(\mathbfcal{I}-\widetilde{\mathbfcal{P}}_\mathrm{as})\wedge\mathbfcal{P}_\mathrm{as}=(\mathbfcal{I}-\widetilde{\mathbfcal{P}}_\mathrm{as}){\mathbfcal{P}}_\mathrm{as}$ since $\mathbfcal{P}_\mathrm{as}$ and $\widetilde{\mathbfcal{P}}_\mathrm{as}$ commute---see Appendix~\ref{app:projectors}.

The spectra of the uncoupled subsystems, shown as colored areas in Fig.~\ref{fig:ps:third}, clearly occupy distinct frequency ranges and reproduce the two peaks of the high-frequency doublet remarkably well. A small discrepancy---the white space between $\rho_\mathrm{as}$ (blue line) and the two colored areas---indicates that the recoupling between $\widetilde{\mathbf{E}}_\mathrm{as}$ and $\widetilde{\mathbf{E}}_\mathrm{as}^\perp$ induces only minimal level repulsion and therefore weak hybridization.

This confirms that the splitting of the high frequency doublet originates from the stiffness contrast between \emph{isotropic} and \emph{deviatoric} tetrahedral stretches~\cite{sarnthein97}, and allows us to precisely identify the subspaces associated with each peak.

\subsubsection{Third level splitting II: the mid-frequency band}
\label{sec:mid}
The mid-frequency band $\mathbf{E}_\mathrm{as}^\perp$ is by definition the $2N/3$-dimensional orthogonal complement of $\mathbf{E}_\mathrm{as}$ within $\mathbf{E}_\mathrm{bs}$. Its elements are polarization vectors $\mathbf{e}\in\mathbf{E}_\mathrm{bs}$ satisfying the $2N/3$ orthogonality constraints $\mathbfcal{A}_\mathrm{as}^\intercal\mathbf{e}=0$. The same counting argument as in the previous section shows that these $2N/3$ constraints cannot be satisfied by elements of $\mathbf{E}_\mathrm{iso}$, so that $\mathbf{E}_\mathrm{as}^\perp\cap\mathbf{E}_\mathrm{iso}=\{\mathbf{0}\}$. However, the subspace $\widetilde{\mathbf{E}_\mathrm{as}^\perp}=\mathbf{E}_\mathrm{as}^\perp\cap\mathbf{E}_\mathrm{dev}$ has dimension $\dim(\widetilde{\mathbf{E}_\mathrm{as}^\perp})=N/3$. Therefore, to capture the stiffness contrast between isotropic and deviatoric tetrahedral stretch contributions, the mid-frequency band must be decomposed into two $N/3$-dimensional subspaces as $\mathbf{E}_\mathrm{as}^\perp=\widetilde{\mathbf{E}_\mathrm{as}^\perp}\oplus\widetilde{\mathbf{E}_\mathrm{as}^\perp}^\perp$.

The spectra of the uncoupled systems in this decomposition are shown as colored areas in Fig.~\ref{fig:ps:third} and clearly occupy distinct, though slightly overlapping, frequency ranges. This decomposition successfully separates the mid-band into two distinct features: $\widetilde{\mathbf{E}_\mathrm{as}^\perp}$ accounts for the sharp peak around \SI{800}{cm^{-1}}, while its complement $\widetilde{\mathbf{E}_\mathrm{as}^\perp}^\perp$ accounts for the shoulder extending over the range $\sim400$--\SI{750}{cm^{-1}}.

Let us further analyze the mode content of the peak near \SI{800}{cm^{-1}}. Its successful capture by $\widetilde{\mathbf{E}_\mathrm{as}^\perp}$ is consistent with its presence in the spectrum of $\mathbf{E}_\mathrm{dev}$ [see Fig.~\ref{fig:tetrahedral}], which already suggested that, while it belongs to $\mathbf{E}_\mathrm{as}^\perp$, it is mainly composed of deviatoric modes. It is noteworthy, however, that this peak is also visible---albeit with in a distorted form---in the spectrum of $\mathbf{E}_\mathrm{ss}$, indicating a close connection with symmetric-stretch modes.

In order to examine this connection, let us consider an arbitrary polarization vector $\mathbf{e}=\mathbfcal{A}\,\boldsymbol{\alpha}\in\mathbf{E}_\mathrm{bs}$ and explicitly write the orthogonality conditions defining $\mathbf{E}_\mathrm{as}^\perp$: $\mathbfcal{A}_\mathrm{as}^\intercal\mathbf{e}=\mathbfcal{A}_\mathrm{as}^\intercal\mathbfcal{A}\,\boldsymbol{\alpha}=0$. Each of these $2N/3$ constraints corresponds to a column of $\mathbfcal{A}_\mathrm{as}$, that is, to a specific oxygen atom labeled $j$, and reads:
\begin{equation}\label{eq:asperp:general}
  \!\!\begin{split}
    &(c_{iji}+c_{jij}-c_{ijk})\,\alpha^{(ij)}
    - (c_{kjk}+c_{jkj}-c_{ijk})\,\alpha^{(kj)}\\
    &+\!\sum_{l\ne j}c_{jil}\,\alpha^{(il)}-\!\sum_{m\ne j}c_{jkm}\,\alpha^{(km)} = 0 \,,
  \end{split}
\end{equation}
where $i$ and $k$ label the silicon atoms bonded to oxygen $j$, and where $l$ (resp. $m$) runs over the oxygen atoms bonded to $i$ (resp. $k$), excluding $j$. The coupling coefficients are defined as $c_{pqr}=c_{rqp}=\vec r_{pq}\cdot\vec r_{rq}/m_q$ for any triplet $(p,q,r)$, where the symmetry follows from the commutativity of the scalar product.

These constraints can be further analyzed by considering that, in view of the high rigidity of the tetrahedra, the \ce{Si-O} bond lengths and the \ce{O-Si-O} angles are narrowly distributed (see Fig.~\ref{fig:model}). As a consequence, three families of coefficients are each sharply peaked around a family-specific value: namely, $c_{iji}=\vec r_{ij}^2/m_j$ and $c_{jij}=\vec r_{ij}^2/m_i$, corresponding to squared bond lengths normalized by masses, and $c_{jij'}$ where $i$ labels a silicon atom and $j$ and $j'$ distinct oxygen atoms bonded to $i$, associated with \ce{O-Si-O} angles. This observation supports the introduction of a \emph{homogeneous-tetrahedron approximation}, in which all coefficients within each of these three families are replaced by their respective peak values. Under this approximation, the above expression simplifies to:
\begin{equation}
  \label{eq:easperp}
  \begin{split}
    (c_{iji}+c_{jij}-c_{jil}-c_{ijk})&\left(\alpha^{(ij)}-\alpha^{(kj)}\right)\\
    &+2c_{jil}\left(T_1^{(i)}-T_1^{(k)}\right)=0\\
  \end{split}
\end{equation}
with $l$ denoting any oxygen bonded to silicon $i$ and $\ne j$.

This expression reveals that the constraints defining the three subspaces
\begin{compactitem}
\item $\mathbf{E}_\mathrm{as}^\perp$, defined by Eq.~\eqref{eq:easperp}
\item $\mathbf{E}_\mathrm{ss}$, defined by the condition $\alpha^{(ij)}=\alpha^{(kj)}$
\item $\mathbf{E}_\mathrm{dev}$, defined by $T_1^{(i)}=0$ for all $i$
\end{compactitem}
are not mutually independent. In particular, for any $\mathbf{e}\in\mathbf{E}_\mathrm{dev}$, the second term of Eq.~\eqref{eq:easperp} vanishes, so that within $\mathbf{E}_\mathrm{dev}$ the defining constraints of $\mathbf{E}_\mathrm{as}^\perp$ and $\mathbf{E}_\mathrm{ss}$ are equivalent. Therefore, within this approximation, $\widetilde{\mathbf{E}_\mathrm{as}^\perp}=\mathbf{E}_\mathrm{as}^\perp\cap\mathbf{E}_\mathrm{dev}$ and $\widetilde{\mathbf{E}}_\mathrm{ss}=\mathbf{E}_\mathrm{ss}\cap\mathbf{E}_\mathrm{dev}$ coincide and both reduce to the triple intersection $\mathbf{E}_\mathrm{as}^\perp\cap\mathbf{E}_\mathrm{dev}\cap\mathbf{E}_\mathrm{ss}$. Independently of this approximation, we previously established that $\dim(\widetilde{\mathbf{E}_\mathrm{as}^\perp})=N/3$ and the same counting arguments yield $\dim(\widetilde{\mathbf{E}}_\mathrm{ss})=N/3$. Within the homogeneous-tetrahedron approximation, these two $N/3$-dimensional subspaces are identical.

We validated this expectation by calculating the spectrum of the restricted vibrational problem on $\widetilde{\mathbf{E}}_\mathrm{ss}$ (not shown) and verifying that it is indistinguishable from that of $\widetilde{\mathbf{E}_\mathrm{as}^\perp}$, shown in Fig.~\ref{fig:ps:third}. It follows that the sharp peak around \SI{800}{cm^{-1}} can be interpreted as being associated with the subspace $\widetilde{\mathbf{E}}_\mathrm{ss}=\mathbf{E}_\mathrm{ss}\cap\mathbf{E}_\mathrm{dev}$, which consists of modes that are simultaneously symmetric-stretch and tetrahedral-deviatoric.\\

In closing this discussion, let us raise two subtle issues. First, the intersection $\mathbf{E}_\mathrm{ss}\cap\mathbf{E}_\mathrm{iso}$ is not $\{\mathbf{0}\}$ as counting rules would suggest. The s.s.\ subspace is defined by the $2N/3$ constraints $\alpha^{(ij)}=\alpha^{(kj)}$, for any oxygen $j$ bonded to silicon atoms $i$ and $k$, while any element of $\mathbf{E}_\mathrm{iso}$ (dimension $N/3$) satisfies $\alpha^{(ij)}=\alpha^{(ij')}=T_1^{(i)}/2$ for any pair $j$ and $j'$ of oxygen atoms bonded to the same silicon atom $i$. All these constraints are simultaneously satisfied by any field in which all $\alpha^{(ij)}$ take the same value. Introducing $\mathbf{A}_\mathrm{uni}=\mathbfcal{A}\,\boldsymbol{\alpha}_\mathrm{uni}$ with $\boldsymbol{\alpha}_\mathrm{uni}=\{1,\ldots,1\}$, we therefore obtain $\mathbf{E}_\mathrm{ss}\cap\mathbf{E}_\mathrm{iso}=\vectspan(\mathbf{A}_\mathrm{uni})$, a one-dimensional space. This particular case arises because both types of constraints enforce equality of $\alpha^{(ij)}$ values across different structural units, and are therefore jointly compatible with the uniform bond-stretch field $\boldsymbol{\alpha}_\mathrm{uni}$.

Such a situation is highly anomalous in disordered systems, where vector fields that are not explicitly interdependent typically span distinct directions in the high-dimensional configuration space, and therefore correspond to independent spanning vectors and constraints, which guarantees the validity of counting rules. To illustrate this point, let us revisit the reason why $\mathbf{E}_\mathrm{as}\cap\mathbf{E}_\mathrm{iso}=\{\mathbf{0}\}$, while by-passing the counting argument. Any element of $\mathbf{E}_\mathrm{iso}$ (dimension $N/3$), satisfies $\alpha^{(ij)}=\alpha^{(ij')}=T_1^{(i)}/2$ for any pair $j$ and $j'$ of oxygen atoms bonded to a same silicon atom $i$. Combined with the a.s.\ constraints, it requires that any pair of tetrahedra sharing an oxygen atom must obey $T_1^{(i)}=-T_1^{(k)}$, where $i$ and $k$ label the silicon atoms. If $T_1^{(i)}\ne0$, this cannot be satisfied along a ring containing an odd number of tetrahedra, and is hence excluded. This argument also shows that a solution involving alternating $T_1^{(i)}$ values could exist in principle---and would only marginally correct the counting rule---but only in a highly non-generic, implausible configuration where all rings are even.

Along the same line, let us reexamine the intersection $\mathbf{E}_\mathrm{as}^\perp\cap\mathbf{E}_\mathrm{iso}$ by substituting $\alpha^{(il)}=\alpha^{(ij)}=T_1^{(i)}/2$ and $\alpha^{(km)}=\alpha^{(kj)}=T_1^{(k)}/2$ in the general condition~\eqref{eq:asperp:general}. This yields $(c_{iji}-c_{ijk}+\sum_l\,c_{jil})\,T_1^{(i)}=(c_{kjk}-c_{ijk}+\sum_m\,c_{jkm})\,T_1^{(k)}$, where the sums run over all the oxygen atoms bonded to silicon $i$ and $k$, respectively. Due to the small distortions arising from disorder, the prefactors multiplying $T_1^{(i)}$ and $T_1^{(k)}$ vary from one pair of tetrahedra to another. The existence of tetrahedral rings then makes it impossible to construct a non-zero field satisfying this set of conditions simultaneously for all oxygen atoms. We thus recover $\mathbf{E}_\mathrm{as}^\perp\cap\mathbf{E}_\mathrm{iso}=\{\mathbf{0}\}$, fully consistently with the dimension counting argument. In the homogeneous tetrahedron approximation, however, the two prefactors become identical and the condition reduces to $T_1^{(i)}=T_1^{(k)}$, which is satisfied by $\mathbf{A}_\mathrm{uni}$. Thus, when eliminating disorder, the intersection becomes $\mathbf{E}_\mathrm{as}^\perp\cap\mathbf{E}_\mathrm{iso}=\vectspan(\mathbf{A}_\mathrm{uni})$. This solution is fragile in the sense that it relies on the exact equality of all the prefactors and is lifted by the small disorder-induced coefficient variations.

The second point we would like to raise concerns the subspace $\mathbf{E}_\mathrm{ss}'=\mathbf{E}_\mathrm{ss}\cap\mathbf{E}_\mathrm{as}^\perp$ of symmetric-stretch modes that are also orthogonal to $\mathbf{E}_\mathrm{as}$. The usual dimensional argument supports that this space reduces to $\{\mathbf{0}\}$. In the homogeneous-tetrahedron approximation, however, a drastic simplification occurs: for any $\mathbf{e}\in\mathbf{E}_\mathrm{ss}'$, the condition $\alpha^{(ij)}=\alpha^{(kj)}$ together with Eq.~\eqref{eq:easperp} reduces to $T_1^{(i)}-T_1^{(k)}=0$. This condition, applied to all pairs of corner-sharing tetrahedra, is satisfied by $\mathbf{A}_\mathrm{uni}$ and by any tetrahedral deviatoric field. It follows that, within this approximation, $\mathbf{E}_\mathrm{ss}'\subset\mathbf{E}_\mathrm{dev}\oplus\vectspan(\mathbf{A}_\mathrm{uni})$, and thus  $\mathbf{E}_\mathrm{ss}'=\mathbf{E}_\mathrm{ss}\cap\mathbf{E}_\mathrm{as}^\perp\cap(\mathbf{E}_\mathrm{dev}\oplus\vectspan(\mathbf{A}_\mathrm{uni}))$. We saw previously that, within the same approximation, the two subspaces $\widetilde{\mathbf{E}_\mathrm{as}^\perp}=\mathbf{E}_\mathrm{as}^\perp\cap\mathbf{E}_\mathrm{dev}$ and $\widetilde{\mathbf{E}}_\mathrm{ss}=\mathbf{E}_\mathrm{ss}\cap\mathbf{E}_\mathrm{dev}$ coincide with the triple intersection $\mathbf{E}_\mathrm{as}^\perp\cap\mathbf{E}_\mathrm{dev}\cap\mathbf{E}_\mathrm{ss}$. We therefore obtain a slightly stronger result, which underlies the previous one: within this approximation, the three subspaces $\mathbf{E}_\mathrm{as}^\perp$, $\mathbf{E}_\mathrm{ss}$, and $\mathbf{E}_\mathrm{dev}\oplus\vectspan(\mathbf{A}_\mathrm{uni})$ share a common intersection of dimension $N/3+1$, reflecting a high degree of mutual compatibility between the corresponding constraints. In this light, the peak around \SI{800}{cm^{-1}} not only consists of modes that are both symmetric-stretch and tetrahedral deviatoric, but also corresponds to the part of $\mathbf{E}_\mathrm{ss}$ which, if not for tetrahedral distortions, would be orthogonal to all antisymmetric bond-stretch modes.

%We verified this analysis by decomposing the mid-band as $\mathbf{E}_\mathrm{as}^\perp=\mathbf{E}_\mathrm{ss}'\oplus\mathbf{E}_\mathrm{ss}^{\prime\,\perp}$ and confirming that the spectra of $\mathbf{E}_\mathrm{ss}'$ and $\mathbf{E}_\mathrm{ss}^{\prime\,\perp}$ (not shown) are indistinguishable from those of $\widetilde{\mathbf{E}_\mathrm{as}^\perp}$ and $\widetilde{\mathbf{E}_\mathrm{as}^\perp}^\perp$, respectively.

\subsection{The no-stretch subspace}
We now turn to analyzing the no-stretch subspace, which contains the low-frequency eigenmodes of the vibrational problem.

An explicit expression for the vectors generating this $5N/3$-dimensional subspace is derived in Appendix~\ref{app:ns}. It is obtained by observing that the displacement vectors of all silicon atoms ($N$ degrees of freedom), can be left unconstrained while introducing the no-stretch conditions via oxygen motion. Indeed, the two no-stretch constraints affecting any oxygen atom (labeled $j$) can be enforced locally by enslaving its displacement in the plane $\{\vec r_{ij},\vec r_{kj}\}$ to that of the two silicon atoms (labeled $i$ and $k$) it is bonded to. As a result, oxygen is only free to move along the third \emph{rocking} direction, $\vec r_{ij}\times\vec r_{kj}$ [see Fig.~\ref{fig:local}-(a)]. With $N$ degrees of freedom from silicon displacements and $2N/3$ from oxygen rocking modes, this correctly accounts for the dimension $5N/3$ of the n.s.\ subspace.

This explicit construction of n.s.\ displacement fields highlights the strong coupling induced by the no-stretch constraints between the motions of silicon and oxygen atoms within $\mathbf{E}_\mathrm{ns}$. It also shows that silicon displacements and oxygen rocking motions generate mutually orthogonal subspaces; however, these are not associated with an evident stiffness contrast and therefore do not lead to a meaningful decomposition of $\mathbf{E}_\mathrm{ns}$ into vibrational subspaces.\\

Since the no-stretch conditions eliminate the stiffness associated with changes in \ce{Si-O} bond lengths, the dominant stiffness within the restricted vibrational problem on $\mathbf{E}_\mathrm{ns}$ arises from the short-range repulsive interactions between oxygen atoms along tetrahedral edges, which are forced into close proximity by their bonding to a common silicon atom. With the \ce{Si-O} bond lengths being fixed, the admissible local deformations that directly mobilize these stiff elements are variations of the angles between \ce{Si-O} bonds, i.e., tetrahedral bending.

Each tetrahedron possesses five independent bending degrees of freedom. With $N/3$ tetrahedra, it yields a total of $5N/3$ internal shape degrees of freedom, which coincides exactly with the dimension of $\mathbf{E}_\mathrm{ns}$. The no-stretch subspace can therefore be fully parametrized by tetrahedral bending coordinates, with all atomic displacements being enslaved to these degrees of freedom by the no-stretch constraints.

As mentioned in Sec.~\ref{sec:local} and illustrated in Fig.~\ref{fig:local}-(b) and Fig.~\ref{fig:tetra}, tetrahedral bending distortions decompose into two irreducible symmetry-invariant subspaces $F_2^\mathrm{b}$ and $E$~\cite{decius1977,taraskin97,shcheblanov16}. The associated displacement fields are explicitly derived in Appendix~\ref{sec:bending} and introduced below.

For each oriented pair of vertices $(j_a,j_b)$, we define the vector
\begin{equation}
\vec b_{j_aj_b}= \frac{1}{r_{ij_a}}\frac{(\vec r_{ij_a}\times\vec r_{ij_b})\times\vec r_{ij_a}}{\|(\vec r_{ij_a}\times\vec r_{ij_b})\times\vec r_{ij_a}\|} \,,
\end{equation}
anchored at vertex $j_a$. This vector lies in the $(\vec r_{ij_a},\vec r_{ij_b})$ plane, is perpendicular to $\vec r_{ij_a}$, and points towards vertex $j_b$. It therefore represents an elementary bending displacement of $j_a$ that reduces the angle between the two bonds. In contrast with previous works~\cite{taraskin97,shcheblanov16}, the vectors $\vec b_{j_aj_b}$ are not normalized to unit length as $\|\vec b_{j_aj_b}\|=1/r_{ij_a}$. This choice allows us to avoid introducing an unnecessary symmetry assumption and thereby to account consistently for possible prior distortions of tetrahedra in the reference configuration.

For any displacement field $\mathbf{u}$, the induced variation of the angle $\theta_{j_aij_b}$ between the two vertices is given by $\delta\theta_{j_aij_b}=-\mathbf{L}_{j_aj_b}^{(i)}\cdot\mathbf{u}$, where $\mathbf{L}_{j_aj_b}^{(i)}$ is the $3N$-dimensional coordinate vector with non-zero components: $\vec L_{j_aj_b\ j_a}^{(i)}=\vec b_{j_aj_b}$ on oxygen $j_a$, $\vec L_{j_aj_b\ j_b}^{(i)}=\vec b_{j_bj_a}$ on oxygen $j_b$, and $\vec L_{j_aj_b\ i}^{(i)}=-\vec b_{j_aj_b}-\vec b_{j_bj_a}$ on the central silicon atom. Accordingly, $\mathbf{L}_{j_aj_b}^{(i)}$ can be interpreted as the displacement field that generates an elementary contraction of angle $\theta_{j_aij_b}$. Up to a sign, $\mathbf{L}_{j_aj_b}^{(i)}$ is the gradient of the angle $\theta_{j_aij_b}$ in configuration space. Because the bond angle is invariant under rigid translations and rotations and bond stretches, its gradient is orthogonal to these motions and represents a purely bending deformation.

The irreducible subspace $F_2^\mathrm{b}$ corresponds to antisymmetric changes of angles between complementary vertex pairs and is spanned by the following elementary displacement vectors:
\begin{equation}\label{eq:F2b}
\begin{split}
  \mathbf{L}_{F_2^\mathrm{b},1}^{(i)}&=\mathbf{L}_{j_1j_2}^{(i)}-\mathbf{L}_{j_3j_4}^{(i)} \, ,\\
  \mathbf{L}_{F_2^\mathrm{b},2}^{(i)}&=\mathbf{L}_{j_1j_3}^{(i)}-\mathbf{L}_{j_2j_4}^{(i)} \, ,\\
    \mathbf{L}_{F_2^\mathrm{b},3}^{(i)}&=\mathbf{L}_{j_1j_4}^{(i)}-\mathbf{L}_{j_2j_3}^{(i)} \, .
\end{split}
\end{equation}
The irreducible subspace $E$ is spanned by the following generating vectors:
\begin{equation}\label{eq:E}
\begin{split}
\mathbf{L}_{E,1}^{(i)}&=2\mathbf{L}_{j_1j_2}^{(i)}+2\mathbf{L}_{j_3j_4}^{(i)}-\mathbf{L}_{j_1j_3}^{(i)}-\mathbf{L}_{j_2j_4}^{(i)}\\
&\qquad\qquad\qquad\qquad-\mathbf{L}_{j_1j_4}^{(i)}-\mathbf{L}_{j_2j_3}^{(i)} \, ,\\
\mathbf{L}_{E,2}^{(i)}&=\mathbf{L}_{j_1j_3}^{(i)}+\mathbf{L}_{j_2j_4}^{(i)}-\mathbf{L}_{j_1j_4}^{(i)}-\mathbf{L}_{j_2j_3}^{(i)} \, .
\end{split}
\end{equation}
Vector $\mathbf{L}_{E,1}^{(i)}$ generates a symmetric contraction of the vertex pairs $(j_1,j_2)$ and $(j_3,j_4)$ accompanied by a compensating opening of the four remaining angles. Vector $\mathbf{L}_{E,2}^{(i)}$ produces an elementary torsion of the tetrahedron, while preserving the internal angles of these vertex pairs. Although these deformation modes have distinct geometric interpretations, they span the same irreducible invariant subspace, since the sum of two torsional modes (of the form $\mathbf{L}_{E,2}^{(i)}$) is a symmetric bending mode (of the form $\mathbf{L}_{E,1}^{(i)}$). Accordingly, $E$-type displacement fields may equivalently be interpreted as describing either tetrahedral symmetric bending or tetrahedral torsion.

Note that the normalization of the generating vectors $\mathbf{L}_{F_2^\mathrm{b},m}^{(i)}$ and $\mathbf{L}_{E,m}^{(i)}$ is irrelevant, since they only serve to define image subspaces and co-kernels. Because these two families of modes span distinct irreducible subspaces, they are expected to be associated with different stiffness scales, and to couple differently to other forms of atomic motion, including the translational and rotational degrees of freedom of tetrahedra.\\

Next, we define the matrices $\mathbfcal{L}_{F_2^\mathrm{b}}$ and $\mathbfcal{L}_{E}$ the columns of which are the vectors $\{\mathbf{L}^{(i)}_{F_2^\mathrm{b},m}, m=1,2,3\}$ and $\{\mathbf{L}^{(i)}_{E,m}, m=1,2\}$, respectively, as well as the corresponding mass-weighted matrices $\mathbfcal{A}_{F_2^\mathrm{b}}=\mathbfcal{M}^{-1/2}\mathbfcal{L}_{F_2^\mathrm{b}}$ and $\mathbfcal{A}_{E}=\mathbfcal{M}^{-1/2}\mathbfcal{L}_{E}$. Following our convention, the associated image subspaces are denoted $\mathbf{E}_{F_2^\mathrm{b}}=\mathrm{Im}(\mathbfcal{A}_{F_2^\mathrm{b}})$ and $\mathbf{E}_E=\mathrm{Im}(\mathbfcal{A}_{E})$. They have dimensions $\dim(\mathbf{E}_{F_2^\mathrm{b}})=N$ and $\dim(\mathbf{E}_E)=2N/3$.

\begin{figure}[t]
  \begin{center}
    \includegraphics[width=0.1\textwidth]{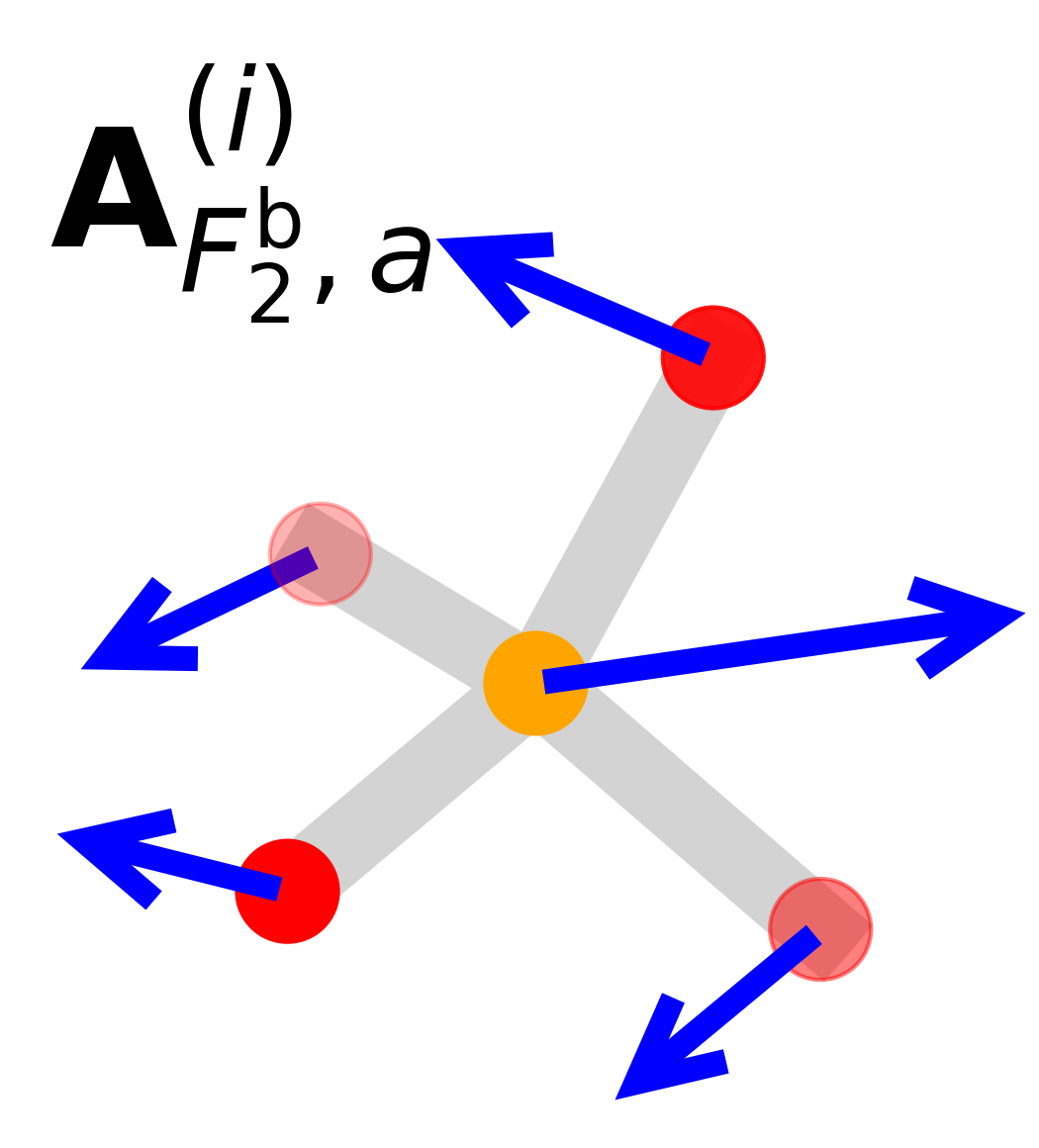}\hfil
    \includegraphics[width=0.1\textwidth]{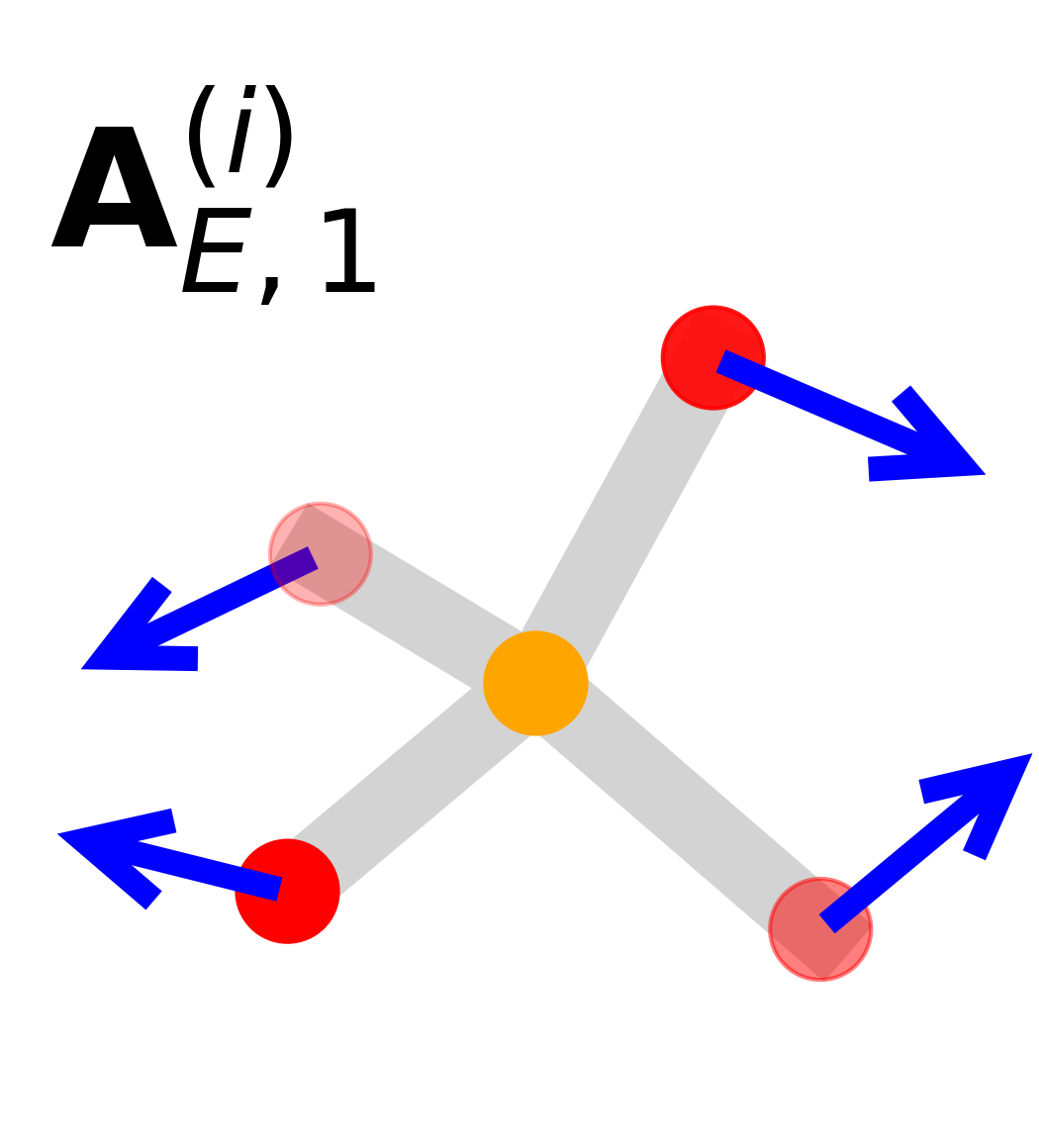}
    \hfil
    \includegraphics[width=0.1\textwidth]{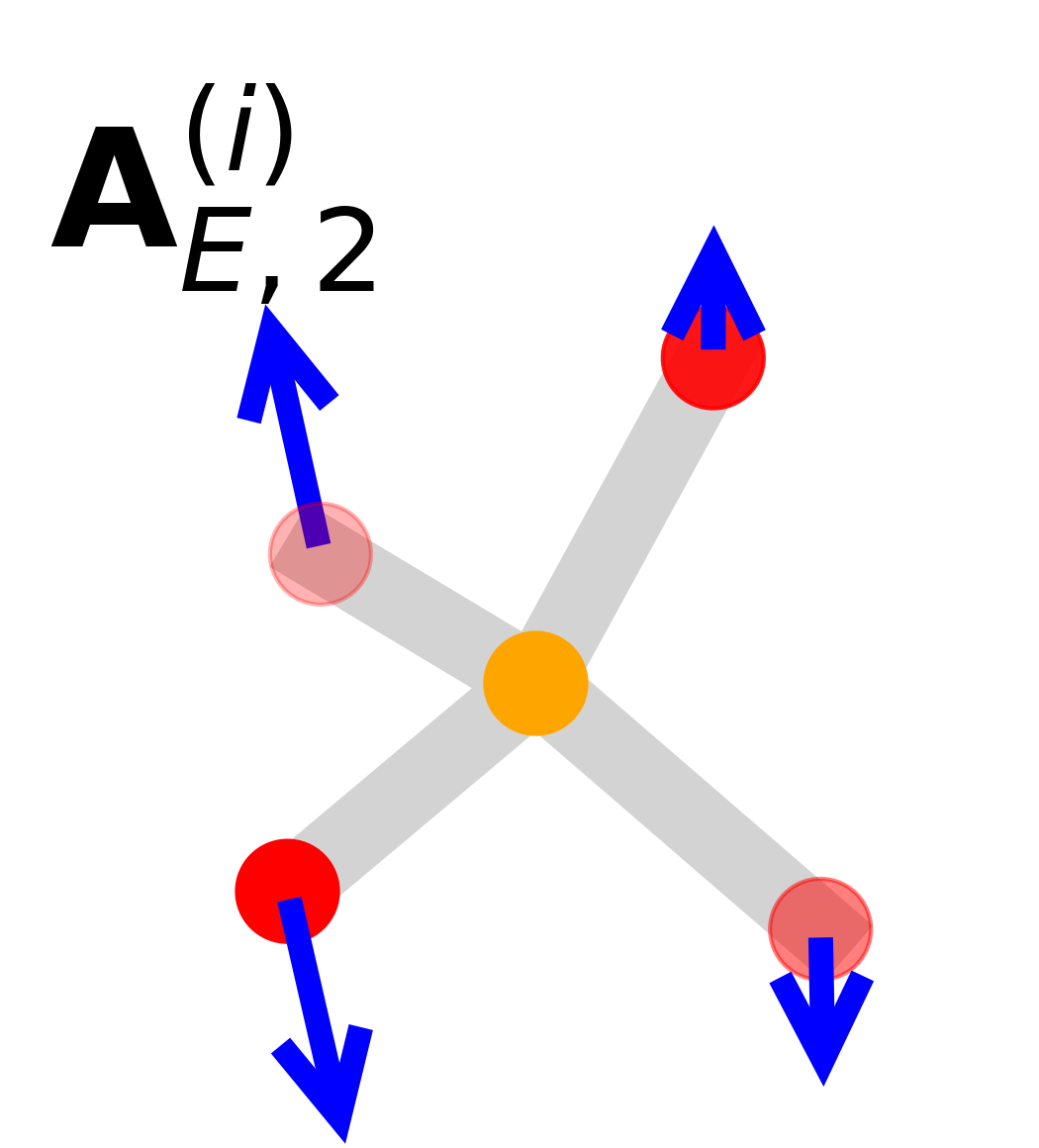}
  \end{center}
  \caption{\label{fig:bending} Bending deformation modes of a tetrahedron as defined in Eqs.~\eqref{eq:F2b} and~\eqref{eq:E}---center in orange, vertices in red---with all atomic contributions to emphasize the different involvement of silicon motion---compare with Fig.~\ref{fig:tetra}. The triply degenerate antisymmetric bending modes ($F_2^{\mathrm{b}}$) exhibits a pronounced antagonistic motion of the silicon atom as all oxygen displacements atoms reinforce one another along the bisector of the bent angles. In contrast, for the symmetric bending and torsional modes ($E$), the silicon displacement vanishes in a regular tetrahedron, as the oxygen motions exactly compensate; in an irregular tetrahedron, a small silicon displacement appears due to pre-existing distortions in the reference configuration.}
\end{figure}

Let us note that, like the elementary bond-stretch displacements of $\mathbfcal{M}^{-1/2}\mathbfcal{A}$ [see the beginning of Sec.~\ref{sec:bs}], the displacement fields associated with the above mass-weighted matrices---i.e., the column vectors of $\mathbfcal{M}^{-1/2}\mathbfcal{A}_{F_2^\mathrm{b}}=\mathbfcal{M}^{-1}\mathbfcal{L}_{F_2^\mathrm{b}}$ and $\mathbfcal{M}^{-1/2}\mathbfcal{A}_{E}=\mathbfcal{M}^{-1}\mathbfcal{L}_{E}$---leave the center of mass unchanged, since the underlying elementary bending displacements $\mathbf{L}_{F_2^\mathrm{b},m}^{(i)}$ and $\mathbf{L}_{E,m}^{(i)}$ are orthogonal to rigid translations. This is a direct consequence of these modes representing purely internal deformations of the considered structural units. The preservation of the center of mass entails an antagonistic motion of the silicon atom illustrated in Fig.~\ref{fig:bending}, where schematic representations of the modes include the silicon component. In a regular tetrahedron, the silicon motion is prominent in $F_2^\mathrm{b}$ modes but vanishes identically by symmetry in the $E$ modes. In the latter modes, it can therefore arise only from pre-existing distortions of the reference configuration.

Although elementary bending modes generate no-stretch deformations of the central tetrahedron (around silicon $i$), they induce bond-length changes in the neighboring tetrahedra sharing the same oxygen atoms. Therefore, the image subspaces of matrices $\mathbfcal{A}_{F_2^\mathrm{b}}$ and  $\mathbfcal{A}_{E}$ are not contained within the n.s.\ subspace. Moreover, since there are $4N/3$ no-stretch constraints, the intersections of either $\mathbf{E}_{F_2^\mathrm{b}}$ or $\mathbf{E}_{E}$ with $\mathbf{E}_\mathrm{ns}$ reduce to the trivial subspace $\{\mathbf{0}\}$. The underlying reason is that $\mathbf{E}_{F_2^\mathrm{b}}$ and $\mathbf{E}_{E}$ correspond to pure linear combinations of bending modes, hence do not include translational and rigid-body rotational motion of tetrahedra, thereby excluding enslaved degrees of freedom that are required to accommodate the no-stretch constraints.

As a result, to separate bending contributions, we must consider the intersections of the co-kernels $\mathbf{E}_{F_2^\mathrm{b}}^\perp$ and $\mathbf{E}_E^\perp$ with $\mathbf{E}_\mathrm{ns}$, to eliminate selected bending components while enforcing no-stretch conditions. The polarization vectors in $\mathbf{E}_{F_2^\mathrm{b}}^\perp$ satisfy $\mathbfcal{A}_{F_2^\mathrm{b}}^\intercal\mathbf{e}=0$ and therefore correspond to displacement fields containing no $F_2^\mathrm{b}$ tetrahedral bending. Accordingly, the subspace $\mathbf{E}_{\mathrm{ns},E}=\mathbf{E}_\mathrm{ns}\cap\mathbf{E}_{F_2^\mathrm{b}}^\perp$ consists of n.s.\ polarization fields containing only tetrahedral $E$ bending components. Its dimension is $\dim(\mathbf{E}_{\mathrm{ns},E})=2N/3$, since $\dim(\mathbf{E}_\mathrm{ns})=5N/3$, while the condition $\mathbfcal{A}_{F_2^\mathrm{b}}^\intercal\mathbf{e}=0$ introduces $N$ independent constraints. Consistently, $\dim(\mathbf{E}_{\mathrm{ns},E})$ is the number of $E$-type bending degrees of freedom. Similarly, $\mathbf{E}_{\mathrm{ns},F_2^\mathrm{b}}=\mathbf{E}_\mathrm{ns}\cap\mathbf{E}_{E}^\perp$ is the $N$-dimensional subspace of $\mathbf{E}_\mathrm{ns}$ parametrized by $F_2^\mathrm{b}$ bending components.

In view of these elements, two decompositions of the n.s.\ subspace can be considered to capture the local stiffness contrast associated with the two types of tetrahedral bending: $\mathbf{E}_\mathrm{ns}=\mathbf{E}_{\mathrm{ns},E}\oplus\mathbf{E}_{\mathrm{ns},E}^\perp$ or $\mathbf{E}_\mathrm{ns}=\mathbf{E}_{\mathrm{ns},F_2^\mathrm{b}}\oplus\mathbf{E}_{\mathrm{ns},F_2^\mathrm{b}}^\perp$. We are therefore led to examine whether either one underlies the double-humped structure of the n.s.\ band.\\

\begin{figure}[!t]
  \includegraphics[width=0.47\textwidth]{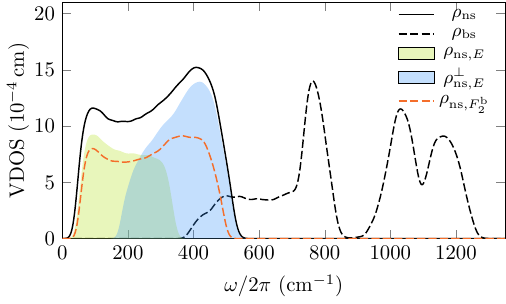}
  \caption{\label{fig:ns:second} Orthogonal decomposition of the n.s.\ subspace as $\mathbf{E}_\mathrm{ns}=\mathbf{E}_{\mathrm{ns},E}\oplus\mathbf{E}_{\mathrm{ns},E}^\perp$, with $\rho_{\mathrm{ns},E}$ and $\rho_{\mathrm{ns},E}^\perp$ denoting the spectra of the uncoupled subspaces $\mathbf{E}_{\mathrm{ns},E}$ and $\mathbf{E}_{\mathrm{ns},E}^\perp$, respectively. For comparison, the spectrum of the n.s.\ subspace, $\rho_\mathrm{ns}$, is shown as a black dashed line, and that of $\mathbf{E}_{\mathrm{ns},F_2^\mathrm{b}}$ as an orange dashed line.}
\end{figure}

The answer to this question is provided in Fig.~\ref{fig:ns:second}, which shows that the spectra of the uncoupled $\mathbf{E}_{\mathrm{ns},E}$ and $\mathbf{E}_{\mathrm{ns},E}^\perp$ capture the upper and lower edges of the n.s.\ spectrum, respectively, and track the corresponding peaks, while occupying distinct, though partially overlapping, frequency ranges. Due to the substantial overlap, the peak in $\rho_{\mathrm{ns},E}$ is somewhat lower than the corresponding one in $\rho_\mathrm{ns}$, indicating a significant level repulsion from recoupling. The spectrum of $\mathbf{E}_{\mathrm{ns},F_2^\mathrm{b}}$, also shown in Fig.~\ref{fig:ns:second}, demonstrates that the alternative decomposition $\mathbf{E}_\mathrm{ns}=\mathbf{E}_{\mathrm{ns},F_2^\mathrm{b}}\oplus\mathbf{E}_{\mathrm{ns},F_2^\mathrm{b}}^\perp$ does not capture the dominant stiffness contrast within the n.s.\ subspace.

The successful decomposition of the low-frequency n.s.\ subspace as $\mathbf{E}_\mathrm{ns}=\mathbf{E}_{\mathrm{ns},E}\oplus\mathbf{E}_{\mathrm{ns},E}^\perp$ follows the same principle as the primary n.s./b.s.\ splitting. Indeed, by defining $\mathbf{E}_{\mathrm{ns},E}=\mathbf{E}_\mathrm{ns}\cap\mathbf{E}_{F_2^\mathrm{b}}^\perp$, we isolate a lower-frequency subspace that is entirely devoid of the stiffer $F_2^\mathrm{b}$ contributions, just as $\mathbf{E}_\mathrm{ns}$ was constructed to eliminate \ce{Si-O} bond stretches. The key technical difference is that here, in contrast to the n.s./b.s.\ splitting, the complementary, higher-frequency subspace $\mathbf{E}_{\mathrm{ns},E}^\perp$ cannot be constructed explicitly by linear combination of the $F_2^\mathrm{b}$ generating vectors, because the subspace they generate, $\mathbf{E}_{F_2^\mathrm{b}}$, is not a subset of $\mathbf{E}_\mathrm{ns}$ and therefore cannot provide a direct parametrization of $\mathbf{E}_{\mathrm{ns},E}^\perp$.

Finally, we conclude that the two-peak structure of the n.s.\ spectrum originates from the stiffness contrast between $E$ (symmetric bending and torsion, softer) and $F_2^\mathrm{b}$ (antisymmetric bending, stiffer) tetrahedral strains. In particular, the lower-frequency part of the vibrational spectrum, which determines acoustic properties, is primarily governed by tetrahedral torsions and symmetric bending ($E$ modes).

This conclusion sharply contrasts with the widespread view that rigid-unit rotations control low-frequency modes~\cite{buchenau86,dove97,nakayama2002}. This idea was motivated by the high stiffness of both \ce{Si-O} bonds and \ce{O-O} tetrahedral edges, which strongly penalize tetrahedral distortions. However, as shown above, rigid-unit motions---tetrahedral translations and rotations---are not independent degrees of freedom but are fully enslaved, through the no-stretch constraints, to the bending coordinates. Consequently, there exists no finite-dimensional subspace of $\mathbf{E}_\mathrm{ns}$ in which tetrahedra undergo purely rigid-unit motion without bending, and therefore no decomposition of $\mathbf{E}_\mathrm{ns}$ that isolates rigid-unit rotations as independent degrees of freedom.

The substantial rotational character observed in the low-frequency modes~\cite{taraskin97,shcheblanov16} (also see Fig.~\ref{fig:local}-(b)) therefore reflects a strong entrainment of rotational motion by the softer $E$ bending modes, due to the kinematic coupling imposed by the no-stretch constraints. It is thus a manifestation of the displacement structure of $E$ modes, rather than the signature of independent rigid-unit rotations associated with a distinct stiffness scale.

%But the no-stretch constraints provide a parametrization of the subspace, but changes in these coordinates impart atomic motion that can be viewed as translations and rotations of the tetrahedra. In our analysis of the b.s.\ subspace, we could construct an explicit representation of the s.s.\ and a.s.\ subspaces in which a.s.\ vectors could be associated with systematically stiffer contributions. The additive combination of such vectors produces the high-frequency doublet.
%It shows that although the stiffness contrast between , they naturally form the high-frequency doublet. Here, the explicit representation of the no-stretch subpace cannot be associated with a stiffness contrast and hence, we cannot construct the parametrization in terms of tetrahedral bending coordinates defines atomic displacements implicitly via the no-stretch conditions.

\section{Discussion \& Conclusion}
We have presented a general framework (ROSA) enabling the decomposition of the vibrational spectrum of amorphous solids. ROSA differs from all prior attempts to analyze the density of states in that it performs a rigorous hierarchical decomposition of the full space of vibrational degrees of freedom through successive applications of the projection formalism. Here, in the example of vitreous silica, the key steps of the analysis were as follows:
\begin{compactitem}
\item A first splitting of the full polarization space based on the rigidity of the \ce{Si-O} bonds, leading to the identification of the no-stretch (n.s.) and bond-stretch (b.s.) subspaces. This primary splitting substantially restricts the admissible local deformation modes at the tetrahedral scale in each subspace: only tetrahedral bending modes, which preserve \ce{Si-O} bond lengths, are compatible with the defining constraints of the n.s.\ subspace; conversely, only tetrahedral stretch modes (isotropic and deviatoric) are produced by linear combinations of elementary bond stretches that span the b.s.\ subspace. Each subspace is fully parametrized by its corresponding set of tetrahedral distortions.
\item In the high-frequency bond-stretch subspace (dimension $4N/3$), the dominant stiffness contrast is associated with the local orientation of oxygen motion relative to the two silicon atoms it is bonded to. The antisymmetric stretch (a.s.) modes are the stiffest ones and constitute the high-frequency doublet $\mathbf{E}_\mathrm{as}$ (dimension $2N/3$), enabling a clear frequency separation through the decomposition $\mathbf{E}_\mathrm{bs}=\mathbf{E}_\mathrm{as}\oplus\mathbf{E}_\mathrm{as}^\perp$. The mid-frequency subspace $\mathbf{E}_\mathrm{as}^\perp$ (dimension $2N/3$) contains all symmetric stretch (s.s.) contributions, but differs from the associated subspace, $\mathbf{E}_\mathrm{ss}$, which is not orthogonal to $\mathbf{E}_\mathrm{as}$.
\item Still in the b.s.\ subspace, a further stiffness contrast exists between isotropic and deviatoric tetrahedral stretches. This contrast, however, is secondary to the a.s./s.s.\ one, hence becomes relevant only when splitting the $2N/3$-dimensional subspaces $\mathbf{E}_\mathrm{as}$ and $\mathbf{E}_\mathrm{as}^\perp$. Since the subspace $\mathbf{E}_\mathrm{iso}$ (dimension $N/3$) generated by isotropic strains is too small to accommodate the $2N/3$ defining constraints of either subspace within $\mathbf{E}_\mathrm{bs}$, the splittings must be constructed by considering their intersections with $\mathbf{E}_\mathrm{dev}$ (dimension $N$). This leads to the identification of four mutually orthogonal $N/3$-dimensional subspaces that generate $\mathbf{E}_\mathrm{bs}$.
\item The low-frequency, no-stretch, subspace (dimension $5N/3$) can be decomposed by separating tetrahedral bending modes as $\mathbf{E}_\mathrm{ns}=\mathbf{E}_{\mathrm{ns},E}\oplus\mathbf{E}_{\mathrm{ns},E}^\perp$. This decomposition is obtained by excluding the stiffest $F_2^\mathrm{b}$ bending contributions through the $N$ orthogonality constraints $\mathbfcal{A}_{F_2^\mathrm{b}}^\intercal\mathbf{e}=0$. It thereby isolates the $2N/3$-dimensional subspace $\mathbf{E}_{\mathrm{ns},E}=\mathbf{E}_\mathrm{ns}\cap\mathbf{E}_{F_2^\mathrm{b}}^\perp$, which only contains $E$ bending contributions and dominates the lower-frequency part of the n.s.\ spectrum.
\end{compactitem}

\begin{figure}[!t]
  \medskip
  \includegraphics[width=0.47\textwidth]{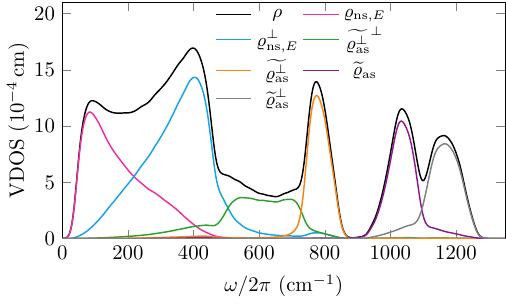}
  \caption{\label{fig:complete} Total and partial VDOS spectra corresponding to the decomposition of the polarization vector space into: $\mathbf{E}=\mathbf{E}_{\mathrm{ns},E}\oplus\mathbf{E}_{\mathrm{ns},E}^\perp\oplus\widetilde{\mathbf{E}_\mathrm{as}^\perp}^\perp\oplus\widetilde{\mathbf{E}_\mathrm{as}^\perp}\oplus\widetilde{\mathbf{E}}_\mathrm{as}\oplus\widetilde{\mathbf{E}}_\mathrm{as}^\perp$.}
\end{figure}

At the end of this procedure, we have obtained a rigorous orthogonal decomposition of the whole polarization vector space into six subspaces:
\begin{equation}
\mathbf{E}=\mathbf{E}_{\mathrm{ns},E}
\oplus\mathbf{E}_{\mathrm{ns},E}^\perp
\oplus\widetilde{\mathbf{E}_\mathrm{as}^\perp}^\perp
\oplus\widetilde{\mathbf{E}_\mathrm{as}^\perp}
\oplus\widetilde{\mathbf{E}}_\mathrm{as}
\oplus\widetilde{\mathbf{E}}_\mathrm{as}^\perp \,.
\end{equation}
In order to assess its relevance to the original vibrational problem, we report in Fig.~\ref{fig:complete} the partial VDOS calculated using Eq.~\eqref{eq:partial}. In contrast with the spectra of the uncoupled vibrational problems considered in the projection method, which guided our analysis and led to this decomposition, these partial VDOS are obtained by decomposing the modes of the \emph{full} (recoupled) vibrational problem into the six orthogonal subspaces. They therefore provide a direct measure of the contribution of each subspace to eigenmodes at a given frequency. The data in Fig.~\ref{fig:complete} show that, in sharp contrast with atomic decompositions or local projection techniques (see Section~\ref{sec:art}), these partial VDOS are strongly concentrated in distinct and ordered frequency ranges and selectively capture all salient features of the spectrum, namely five peaks and an intermediate low-density region between \SI{500}{cm^{-1}} and \SI{700}{cm^{-1}}. This establishes a precise and rigorous identification of the dominant classes of atomic motion across the vibrational spectrum, which constitutes a major outcome of our work.

Hybridization effects remain moderate. The high-frequency doublet does not hybridize with the rest of the spectrum, but a weak hybridization is visible between the $\widetilde{\mathbf{E}}_\mathrm{as}$ and $\widetilde{\mathbf{E}}_\mathrm{as}^\perp$ subspaces. In the peak near \SI{800}{cm^{-1}}, the dominant $\widetilde{\mathbf{E}_\mathrm{as}^\perp}$ modes (shown in Sec.~\ref{sec:mid} to correspond to tetrahedral deviatoric and symmetric-stretches) exhibit weak hybridization with the $\widetilde{\mathbf{E}_\mathrm{as}^\perp}^\perp$ modes (lower-part of the mid-frequency band) and, to a lesser extent, with modes from the high-frequency part of the no-stretch subspace $\mathbf{E}_{\mathrm{ns},E}^\perp$. Modes belonging to $\widetilde{\mathbf{E}_\mathrm{as}^\perp}^\perp$ are found down to \SI{200}{cm^{-1}}, which indicates some degree of hybridization with $\mathbf{E}_{\mathrm{ns},E}^\perp$. Hybridization between the two no-stretch subspaces ($\mathbf{E}_{\mathrm{ns},E}$ and $\mathbf{E}_{\mathrm{ns},E}^\perp$) is evident from the comparison between partial VDOS and the spectra of the corresponding restricted problems in Fig.~\ref{fig:ns:second}. It is most pronounced in the frequency region where $\rho_{\mathrm{ns},E}$ and $\rho_{\mathrm{ns},E}^\perp$ overlap, indicating that it occurs primarily via a resonant mechanism.\\

A subtle feature of any decomposition of the vibrational space is that two types of local displacement fields arise, depending on whether they are used to define co-kernels (constraints) or image subspaces.

Since constraints are meant to suppress local deformation modes, they must be introduced through the configuration-space gradients of local shape parameters~\cite{decius1977,hamermesh1962,herzberg1991}---such as bond lengths, tetrahedral stretches or angles. By definition, these shape parameters are invariant under rigid translations, and their gradients are therefore orthogonal to translations of the corresponding structural units. The resulting displacement fields, which are the columns of the matrices we systematically denote by the symbol $\mathbfcal{L}$ with appropriate indices, are therefore translation-free.

Since the splitting is performed in polarization space, these constraints must be expressed through the mass-weighted matrix $\mathbfcal{A}=\mathbfcal{M}^{-1/2}\mathbfcal{L}$. Although this relation results from a simple change of variable, it has an important consequence: the image subspace of $\mathbfcal{A}$ is spanned by displacement vectors given by the columns of $\mathbfcal{M}^{-1/2}\mathbfcal{A}=\mathbfcal{M}^{-1}\mathbfcal{L}$. These vectors are not translation-free but instead preserve the center of mass. Thus, while the local displacement fields that introduce constraints are translation-free, those that span image subspaces preserve the center of mass.

The most immediate physical consequence of this algebraic structure appears in the first splitting of the polarization subspace into the no-stretch and bond-stretch components. Because the bond-stretch subspace is the image of matrix $\mathbfcal{A}$, it is spanned by displacement fields that preserve the center of mass; their kinetic energy therefore averages out beyond the bond scale. Since plane waves mobilize the collective inertia of extended groups of atoms at the scale of their wavelength, we expect their projection onto bond-stretch modes to vanish in the large-wavelength limit.

In this light, the splitting between the no-stretch and bond-stretch subspace is reminiscent of the distinction between acoustic and optic modes in crystals. Long-wavelength phonons should primarily couple to no-stretch modes, as they mobilize the inertia of extended atomic domains. By contrast, bond-stretch modes involve strongly local and antagonistic displacements of bonded atoms and therefore should couple only weakly to long-wavelength phonons, in a manner reminiscent of optic modes in crystals. A thorough elaboration of this analogy lies beyond the scope of the present work, but constitutes a fundamental issue deserving further investigation.\\

Let us emphasize the key role of global orthogonality constraints in shaping the spectral structure. This appears more clearly when analyzing the bond-stretch subspace: the high-frequency doublet is then directly determined by the intrinsic frequencies of the antisymmetric-stretch subspace ($\mathbf{E}_\mathrm{as}$), while the symmetric-stretch subspace ($\mathbf{E}_\mathrm{ss}$), its natural complement at the scale of bonded \ce{Si-O-Si} units, displays lower intrinsic frequencies largely overlapping with the mid-frequency range and account for the \SI{800}{cm^{-1}} peak. However, although $\mathbf{E}_\mathrm{ss}$ and $\mathbf{E}_\mathrm{as}$ together generate the entire b.s.\ subspace, they are not mutually orthogonal. Consequently, $\mathbf{E}_\mathrm{as}^\perp$ can be viewed as the orthogonal projection of $\mathbf{E}_\mathrm{ss}$ away from $\mathbf{E}_\mathrm{as}$. As explained at the end of Sec.~\ref{sec:asss}, this projection produces modes whose frequencies are systematically shifted downward (away from those of the a.s.\ doublet) compared with intrinsic s.s.\ ones. As a result, the $\mathbf{E}_{\mathrm{as}}^\perp$ spectrum lies at systematically lower frequencies than that of $\mathbf{E}_{\mathrm{ss}}$, leading, in particular, to the development of the low-frequency shoulder of the mid-frequency band.

The same phenomenon should in principle take place for any splitting, but in other cases we did not systematically identify non-orthogonal generating subspaces.

Let us also point out that, while some subspaces admit an explicit interpretation in terms of displacement fields constructed from specific types of local contributions, others can only be defined implicitly through global orthogonality constraints. Their physical interpretation therefore cannot be based on a direct construction in terms of elementary motions and their associated stiffnesses, but follows instead from the exclusion of contributions with contrasting stiffnesses enforced by these constraints.

For example, when splitting $\mathbf{E}_\mathrm{as}$ based on tetrahedral stretches, the lower part of the high-frequency doublet ($\widetilde{\mathbf{E}}_\mathrm{as}$) consists of all polarization vectors involving only antisymmetric and tetrahedral deviatoric stretches---which provides an immediate interpretation of the nature of this subspace. By contrast, $\widetilde{\mathbf{E}}_\mathrm{as}^\perp$, its complement in $\mathbf{E}_\mathrm{as}$, combines antisymmetric stretches with both deviatoric and isotropic tetrahedral contributions. It therefore cannot be characterized by the presence of a specific type of elementary tetrahedral contribution, but must be understood through the orthogonality constraints that define it---the exclusion of the purely deviatoric antisymmetric-stretch modes of $\widetilde{\mathbf{E}}_\mathrm{as}$.

Similarly, the peak around $\SI{800}{cm^{-1}}$ ($\widetilde{\mathbf{E}_\mathrm{as}^\perp}\simeq\widetilde{\mathbf{E}}_\mathrm{ss}=\mathbf{E}_\mathrm{ss}\cap\mathbf{E}_\mathrm{dev}$ [see Sec.~\ref{sec:mid}]) can be identified as corresponding to modes that involve only symmetric and deviatoric tetrahedral stretch contributions. However, its complement, $\widetilde{\mathbf{E}_\mathrm{as}^\perp}^\perp$, which spans the lower part of the mid-frequency band, combines symmetric and antisymmetric stretch as well as deviatoric and isotropic tetrahedral stretch contributions. It therefore cannot be interpreted in terms of a specific class of elementary stretches, but is instead defined implicitly through the orthogonality constraints that exclude the contributions from $\widetilde{\mathbf{E}_\mathrm{as}^\perp}$. More precisely, within the homogeneous tetrahedron approximation $\widetilde{\mathbf{E}_\mathrm{as}^\perp}\simeq\mathbf{E}_\mathrm{ss}\cap\mathbf{E}_\mathrm{dev}$ is the subspace of symmetric-stretch modes that also satisfy the orthogonality constraints defining $\mathbf{E}_\mathrm{as}^\perp$ [Sec.~\ref{sec:mid}]. Therefore, its complement $\widetilde{\mathbf{E}_\mathrm{as}^\perp}^\perp$ necessarily gathers all the contributions from s.s.\ modes that do not satisfy these constraints and are projected away from $\mathbf{E}_\mathrm{as}$, causing a downward frequency shift.

Finally, when splitting the low-frequency no-stretch subspace as $\mathbf{E}_\mathrm{ns}=\mathbf{E}_{\mathrm{ns},E}\oplus\mathbf{E}_{\mathrm{ns},E}^\perp$, the subspace $\mathbf{E}_{\mathrm{ns},E}$ consists of no-stretch modes that involve only $E$ tetrahedral bending. Its complement, $\mathbf{E}_{\mathrm{ns},E}^\perp$, is defined solely through orthogonality constraints and therefore cannot be identified as a subspace generated only by some subset of bending modes. Instead, it comprises no-stretch modes that combine both $E$ and $F_2^\mathrm{b}$ bending contributions, which remaining orthogonal to the pure-$E$ bending subspace $\mathbf{E}_{\mathrm{ns},E}$.\\

Having obtained a rigorous orthogonal decomposition of the polarization space that captures the main features of the vibrational spectrum, we can now clarify several longstanding issues concerning the role of specific local units and symmetry groups in shaping the vibrational spectrum of vitreous silica.

In particular, the literature has so far struggled to determine under which conditions it is meaningful to isolate different atomic contributions---such as silicon motion or oxygen bending, stretching, rocking components---and to understand why these contributions appear throughout the spectrum with varying amplitudes [see Fig.~\ref{fig:local}-(a)]. Our analysis sheds light on this issue by precisely identifying the subspaces in which these different types of motion are admissible and how they couple with other degrees of freedom. In particular, rocking motion is strictly confined to the no-stretch subspace, which explains both its prominence in the corresponding frequency range and its rapid decay above $\simeq\SI{550}{cm^{-1}}$, where it can only appear through weak hybridization with bond-stretch modes.

We also found that there exists no relevant subspace in which oxygen bending and stretching appear as independent contributions. In the n.s.\ subspace, these motions do occur, but are entirely enslaved to silicon displacements, as demonstrated in Appendix~\ref{app:ns}. Moreover, the associated stiffness contrast becomes relevant only in the bond-stretch subspace, where oxygen bending and stretching are necessarily coupled with antagonistic displacements of the bonded silicon atoms. The relevant contrast is therefore that between symmetric and antisymmetric modes, which governs the main splitting of the bond-stretch subspace. This explains why oxygen bending is found (albeit with a small amplitude) in the mid-frequency region, whereas oxygen stretch is found in the high-frequency doublet. The strong confinement of these contributions within their respective frequency ranges---namely, the near absence of oxygen bending in the doublet and of oxygen stretching in the mid-frequency band (except at lower frequencies, presumably due to hybridization with no-stretch modes)---can furthermore be understood from the absence of hybridization between $\mathbf{E}_\mathrm{as}$ and $\mathbf{E}_\mathrm{as}^\perp$ modes.

The coupling with silicon motion, which is the defining feature of the bond-stretch subspace, causes the relative amplitudes of oxygen and silicon displacements to be strongly determined by the relevant local bond-stretch contributions and by the geometry of the associated structural units. For instance, the moderate amplitude of silicon contributions in the high-frequency doublet originates from the geometry of the a.s.\ stretch modes at the scale of a \ce{Si-O-Si} bridge (see Fig.~\ref{fig:asssmodes}). In these modes, the oxygen displacement is enhanced by the positive superposition of oxygen contributions and by the obtuse angle, which entails that the selected component is the largest one. By contrast, the very large relative amplitude of silicon displacements in the peak near \SI{800}{cm^{-1}} cannot be explained by the geometry of s.s.\ local contributions, for which oxygen displacement remains dominant (see Fig.~\ref{fig:asssmodes}). Instead, it should arise because these modes are also tetrahedral deviatoric (see inset of Fig.~\ref{fig:tetrahedral}), which produces a coherent superposition of stretch contributions from the four \ce{Si-O} bonds of each tetrahedron, thereby amplifying the silicon displacement.

Our analysis also clarifies the role of tetrahedral symmetry-adapted modes. In previous studies, ambiguity persisted regarding both their correct definition and spectral relevance. In particular, it remained unclear whether such modes should be defined as translation-free (by considering displacements relative to the silicon)~\cite{taraskin97,shcheblanov16,oligschleger99,mukhopadhyay2003}, or while imposing a fixed center of mass for the whole tetrahedron~\cite{sarnthein97,pasquarello98a}, and how these alternative definitions relate to the different regions of the vibrational spectrum. Our construction shows that this ambiguity reflects the distinct roles played by constraint modes, which are translation-free, and spanning modes, which preserve the center of mass.

Moreover, our analysis clearly shows that the tetrahedral strains that change \ce{Si-O} bond lengths ($A_1$ and $F_2^\mathrm{s}$ symmetries) are necessarily confined to the bond-stretch subspace. Consequently, they are defined for a fixed center of mass and correspond to the isotropic and deviatoric tetrahedral stretch contributions. We have demonstrated that, within this subspace, the stiffness contrast between isotropic and deviatoric stretches is secondary to the dominant contrast between symmetric and antisymmetric stretches. It therefore does not control the primary band separation but only contributes to a third-level splitting of this band, determining the internal structure of the high-frequency doublet and the subdivision of the mid-frequency band into the \SI{800}{cm^{-1}} peak and the lower frequency shoulder (see Fig.~\ref{fig:complete}).

Our conclusion on the role of tetrahedral stretch modes in the splitting of the high-frequency doublet fully agrees with the work of Sarnthein, Pasquarello, and Car~\cite{sarnthein97}, who showed that deviatoric ($T_2$, in their nomenclature) tetrahedral stretch modes predominantly contribute to the lower-frequency part of the high-frequency doublet. Their work was instrumental in establishing that the formation of the doublet is an intrinsic structural feature rather than a consequence of Coulombic TO-LO splitting. However, they did not provide a clear justification as to why tetrahedral modes should be defined while preserving the center of mass; their analysis also found significant deviatoric contributions to the \SI{800}{cm^{-1}} peak without explaining their origin and without recognizing that the isotropic/deviatoric contrast is secondary to the symmetric/antisymmetric stretch contrast.

Meanwhile, tetrahedral bending modes ($F_2^\mathrm{b}$ and $E$), which preserve \ce{Si-O} bond lengths, are strictly confined to the no-stretch subspace and fully parametrize its $5N/3$ degrees of freedom. The stiffness contrast between these two classes of bending modes produces the two-peak structure of the no-stretch spectrum. Importantly, neither silicon translations nor tetrahedral rotations constitute fixed or independent degrees of freedom within this subspace, as both are enslaved to the bending coordinates by the no-stretch constraints.

It should be emphasized that there exists no finite-dimensional subspace in which tetrahedra undergo pure rigid-unit rotational motion without bending. Indeed, treating tetrahedra as rigid units would impose $3N$ independent constraints---$4N/3$ fixing bond lengths and $5N/3$ fixing tetrahedral angles---which together equal the dimension of the entire polarization space. This rules out, on purely geometric grounds, the widespread interpretation that low-frequency modes, including those associated with the Boson peak~\cite{buchenau86,dove97,nakayama2002}, correspond to rigid-unit rotations.

The above results show that the substantial rotational character of low-frequency modes must instead be understood as a consequence of the kinematic coupling between rotations and $E$ tetrahedral bending modes imposed by the no-stretch constraints. Rotational motion is therefore not an independent degree of freedom governing a specific spectral feature or introducing a distinct stiffness scale, but a dependent component of bending modes that dominate the low-frequency range. Accordingly, the stiffness associated with $E$ bending modes (symmetric bending or torsion) must be regarded as a key factor governing the low-frequency part of the spectrum, down into the acoustic range.\\

To conclude, the present work introduces a systematic and physically transparent framework---ROSA---for analyzing the vibrational spectra of covalently bonded disordered solids. ROSA consists in performing repeated orthogonal decompositions of embedded polarization subspaces, each exploiting a dominant stiffness contrast. It thus recursively separates distinct families of local deformation modes and unveils a hierarchy of mutually orthogonal subspaces associated with distinct stiffness scales, thereby enabling a precise structural and modal assignment of vibrational spectral features.

Any tetrahedral glass, such as \ce{GeO2}, \ce{GeSe2}, \ce{GeS2}, \ce{ZnCl2}, can be straightforwardly analyzed via ROSA by considering the same subspaces and determining the role of the associated stiffness contrasts and how they combine with inertial effects arising from different atomic masses. These effects should play a minor role in the bond-stretch subspace, where geometric (no-stretch) constraints tightly couple the motion of bonded atomic pairs and enforce strict preservation of the center of mass for any elementary stretch field. In the no-stretch subspace, by contrast, inertial effects should interplay more strongly with local bending modes, which directly couple to translational and rotational motion.

More generally, because it relies solely on the existence of distinct stiffness scales associated with symmetry-adapted deformation modes of local structural units, and proceeds by introducing geometric constraints via the projection formalism, ROSA is broadly applicable to most covalent network solids. Our work thus opens the way toward a unified structural interpretation of vibrational spectra across a broad class of amorphous materials.

%\ns{In crystalline solids, the vibrational mode assignment follows naturally from translational periodicity and point-group symmetries; in amorphous materials, no analogous guiding principle has been available. In our approach, the role of symmetry is replaced by the hierarchy of stiffness contrasts between distinct classes of atomic motion: at each level, the stiffest contributions are identified and isolated through orthogonal splitting, yielding eigenspaces associated with distinct frequency ranges without relying on any assumption of periodicity.}
%\ns{In particular, the precise identification of orthogonal subspaces and their associated frequency ranges provides a rigorous foundation for interpreting spectroscopic data---such as Raman and infrared spectra---in terms of well-defined classes of atomic motion, moving beyond the qualitative assignments that have prevailed in the literature.}

\begin{acknowledgments}
This work has benefited from the support of the project ViSIONs ANR-18-CE08-0023-03 of the French National Research Agency (ANR), and from a French government grant managed by ANR within the framework of the national program Investments for the Future ANR-11-LABX-022-01 and was provided with computing HPC and storage resources by GENCI at TGCC thanks to the grants AD010914164 on the supercomputer Joliot Curie's SKL \& ROME partitions.

We gratefully acknowledge Mikhail Povarnitsyn for his assistance with the production of $a$-\ce{SiO2} configurations and initial implementation of analysis routines at an early stage of this work.
\end{acknowledgments}

\appendix

\section{Numerical calculation of projectors}
\label{app:projectors}

\subsection{Projectors on image subspace and co-kernel}
The general expression for the projector on the image subspace $\mathbf{E}_\mathrm{A}=\mathrm{Im}(\mathbfcal{A})$ of an $m\times n$ matrix is $\mathbfcal{P}_\mathrm{A}=\mathbfcal{A}\mathbfcal{A}^+$, with $\mathbfcal{A}^+$ an $n\times m$ matrix called the Moore-Penrose \emph{pseudo-inverse}. The projector onto the complement $\mathbf{E}_\mathrm{A}^\perp=\mathrm{Ker}(\mathbfcal{A}^\intercal)$ is just $\mathbfcal{P}_\mathrm{A}^\perp=\mathbfcal{I}-\mathbfcal{P}_\mathrm{A}$.

The pseudo-inverse is the only $n\times m$ matrix that satisfies the four relations: (i) $\mathbfcal{A}\mathbfcal{A}^+\mathbfcal{A}=\mathbfcal{A}$, (ii) $\mathbfcal{A}^+\mathbfcal{A}\mathbfcal{A}^+=\mathbfcal{A}^+$, (iii) $(\mathbfcal{A}\mathbfcal{A}^+)^\intercal=\mathbfcal{A}\mathbfcal{A}^+$, and (iv) $(\mathbfcal{A}^+\mathbfcal{A})^\intercal=\mathbfcal{A}^+\mathbfcal{A}$. In practice, it can be calculated as follows. First, a singular-value decomposition (SVD) of $\mathbfcal{A}$ is performed, i.e., the matrix is written as:
$\mathbfcal{A}=\mathbfcal{U}\mathbfcal{S}\mathbfcal{V}^\intercal$, where $\mathbfcal{U}$ ($m\times m$) and $\mathbfcal{V}$ ($n\times n$) are orthogonal square matrices and $\mathbfcal{S}$ is diagonal, although non-square, of the same dimensions, $m\times n$, as $\mathbfcal{A}$.
%old{$\mathbfcal{A}=\mathbfcal{U}\mathbfcal{S}\mathbfcal{V}^\intercal$, where $\mathbfcal{U}$ and $\mathbfcal{V}$ are orthogonal (hence square) and $\mathbfcal{S}$ is diagonal, although non-square, of the same dimensions, $m\times n$, as $\mathbfcal{A}$.}
The pseudo-inverse is the matrix $\mathbfcal{A}^+=\mathbfcal{V}\mathbfcal{S}^+\mathbfcal{U}^\intercal$, where $\mathbfcal{S}^+$ is the matrix obtained from $\mathbfcal{S}$ by (a) taking the transpose and (b) inverting all the non-zero diagonal elements.

When $\mathbfcal{A}$ is full column rank, $\mathbfcal{A}^\intercal\mathbfcal{A}$ is invertible~\cite{meyer2023} and the pseudoinverse reads: $\mathbfcal{A}^+=\left(\mathbfcal{A}^\intercal\mathbfcal{A}\right)^{-1}\mathbfcal{A}^\intercal$. This can be checked by considering an arbitrary polarization vector $\mathbf{e}=\mathbf{e}_\mathrm{A}+\mathbf{e}_\mathrm{A}^\perp$ with $\mathbf{e}_\mathrm{A}\in\mathbf{E}_\mathrm{A}$ and $\mathbf{e}_\mathrm{A}^\perp\in\mathbf{E}_\mathrm{A}^\perp$. Writing $\mathbf{e}_\mathrm{A}\equiv\mathbfcal{A}\,\boldsymbol{\alpha}$, and applying $\mathbfcal{A}^\intercal$ to the left of this equation, we find $\boldsymbol{\alpha}=(\mathbfcal{A}^\intercal\mathbfcal{A})^{-1}\mathbfcal{A}^\intercal\mathbf{e}=\mathbfcal{A}^+\mathbf{e}$ (using $\mathbfcal{A}^\intercal\mathbf{e}_\mathrm{A}^\perp=0$), and thus $\mathbf{e}_\mathrm{A}=\mathbfcal{A}\mathbfcal{A}^+\mathbf{e}$, so that $\mathbfcal{P}_\mathrm{A}\equiv\mathbfcal{A}\mathbfcal{A}^+$, as announced.

\subsection{Projector on an intersection subspace}
Let us consider two subspaces $\mathbf{G}$ and $\mathbf{H}$, about which we know the associated orthogonal projections $\mathbfcal{P}_\mathrm{G}$ and $\mathbfcal{P}_\mathrm{H}$. The orthogonal projection onto the intersection subspace $\mathbf{G}\cap\mathbf{H}$, denoted $\mathbfcal{P}_\mathrm{G}\wedge\mathbfcal{P}_\mathrm{H}=\mathbfcal{P}_\mathrm{H}\wedge\mathbfcal{P}_\mathrm{G}$, is the large $n$ limit of $(\mathbfcal{P}_\mathrm{H}\mathbfcal{P}_\mathrm{G})^n$ or equivalently $(\mathbfcal{P}_\mathrm{G}\mathbfcal{P}_\mathrm{H})^n$~\cite{BenIsrael2003}.

Its explicit calculation consists in first performing an singular-value decomposition of the composed projector: $\mathbfcal{P}_\mathrm{G}\mathbfcal{P}_\mathrm{H}=\mathbfcal{USV}^\intercal$, or alternatively of $\mathbfcal{P}_\mathrm{H}\mathbfcal{P}_\mathrm{G}$. Then we construct the matrix $\mathbfcal{U}'=\{\mathbf{u}_n\}$ from the columns $\mathbf{u}_n$ of the matrix $\mathbfcal{U}$ corresponding to the singular elements $s_{nn}=1$ of the diagonal matrix $\mathbfcal{S}$; finally, the projector on the intersection subspace is calculated as $\mathbfcal{P}_\mathrm{G}\wedge\mathbfcal{P}_\mathrm{H}=\mathbfcal{U}'\mathbfcal{U'}^\intercal$. In practice, the numerically obtained values $s_{nn}$ are never strictly equal to 1: the dimension of the intersection subspace (number of relevant columns) is first deduced from the constraint counting, and we inspect the elements of $\mathbfcal{S}$ to systematically check that the expected number of values is close to 1 up to numerical errors.

Since $\mathbf{G}\cap\mathbf{H}$ is a subset of both $\mathbf{G}$ and $\mathbf{H}$, its orthogonal projector $\mathbfcal{P}_\mathrm{G}\wedge\mathbfcal{P}_\mathrm{H}$ commutes with both $\mathbfcal{P}_\mathrm{G}$ and $\mathbfcal{P}_\mathrm{H}$. Moreover, whenever $\mathbfcal{P}_\mathrm{G}$ and $\mathbfcal{P}_\mathrm{H}$ commute, we have $\mathbfcal{P}_\mathrm{G}\wedge\mathbfcal{P}_\mathrm{H}=\mathbfcal{P}_\mathrm{G}\mathbfcal{P}_\mathrm{H}=\mathbfcal{P}_\mathrm{H}\mathbfcal{P}_\mathrm{G}$. When either subspace is included in the other, $\mathbfcal{P}_\mathrm{G}\wedge\mathbfcal{P}_\mathrm{H}$ further reduces to the orthogonal projection onto the smaller subspace.

\section{Explicit form of the n.s.\ subspace}
\label{app:ns}
Consider an oxygen atom labeled $j$ and bonded to two silicon atoms labeled $i$ and $k$. In the basis $\{\vec n_{ij},\vec n_{kj},\vec n_{ij}\times\vec n_{kj}\}$, with $\vec n_{ab}=\vec r_{ab}/\|\vec r_{ab}\|$ the normal vector along any \ce{Si-O} bond, and $\times$ denoting the cross product, the displacement vector of oxygen reads:
\begin{equation}
\vec u_j=a_i\,\vec n_{ij}+a_k\,\vec n_{kj}+b\,\vec n_{ij}\times\vec n_{kj}\qquad.
\end{equation}
Using this expression, the no-stretch constraints, $\vec r_{ij}\cdot\vec u_{ij}$ and $\vec r_{kj}\cdot\vec u_{kj}$, can be recast as a $2\times2$ linear problem on $a_i$ and $a_k$:
\begin{equation}
\begin{cases}
a_i+a_k\cos\theta_{ijk}=\vec u_i\cdot\vec n_{ij}\\
a_i\cos\theta_{ijk}+a_k=\vec u_k\cdot\vec n_{kj}
\end{cases} \,\, ,
\end{equation}
with $\theta_{ijk}$ being the \ce{Si-O-Si} angle. The problem is solved as:
\begin{equation}
 \!\!\!\!\!\!\!\! \begin{pmatrix}a_i\\a_k\end{pmatrix}=
  \frac{1}{\sin^2\theta_{ijk}}\!
  \begin{pmatrix}
    1&-\cos\theta_{ijk}\\
    -\cos\theta_{ijk}&1\\
  \end{pmatrix} \!\!
  \begin{pmatrix}
    \vec u_i\cdot\vec n_{ij}\\
    \vec u_k\cdot\vec n_{kj}
    \end{pmatrix}.
\end{equation}
Finally, the displacement vector of the oxygen atom reads:
\begin{equation}
  \begin{split}
    \vec u_j&=\frac{\vec n_{ij}\otimes\vec n_{ij}-\cos\theta_{ijk}\,\vec n_{kj}\otimes\vec n_{ij}}{\sin^2\theta_{ijk}}\cdot\vec u_i\\
    &+\frac{\vec n_{kj}\otimes\vec n_{kj}-\cos\theta_{ijk}\,\vec n_{ij}\otimes\vec n_{kj}}{\sin^2\theta_{ijk}}\cdot\vec u_k\\
    &+b\,\vec n_{ij}\times\vec n_{kj} \, ,
  \end{split}
  \label{eq:ns_oxygen_displacement}
\end{equation}
where $\vec a\otimes\vec b=\vec a\,\vec b^\intercal$ denotes the tensor product, so that $(\vec a\otimes\vec b)\cdot\vec c=\vec a\,(\vec b\cdot\vec c)$.

The matrix $\mathbfcal{S}_\mathrm{ns}$ generating such displacement fields can be constructed as follows. For each silicon atom $i$, it comprises a diagonal block equal to the $3\times3$ identity, accounting for the fact that the displacement vector of this atom is arbitrary; on the same 3 columns, it also contains four $3\times3$ blocks corresponding to the operators $(\vec n_{ij}\otimes\vec n_{ij}-\cos\theta_{ijk}\vec n_{kj}\otimes\vec n_{ij})/\sin^2\theta_{ijk}$ at the locations corresponding to the displacement vectors of the four oxygen atoms bonded to silicon $i$. Additionally, this matrix contains one column for each oxygen, corresponding to the vector $\vec n_{ij}\times\vec n_{kj}$, accounting for the fact that the rocking component is independent. The associated mass-weighted matrix with image in the polarization vector space is $\mathbfcal{A}_\mathrm{ns}=\mathbfcal{M}^{1/2}\mathbfcal{S}_\mathrm{ns}$ and thus $\mathbf{E}_\mathrm{ns}=\mathrm{Im}\,(\mathbfcal{A}_\mathrm{ns})$. Note that the column vectors of $\mathbfcal{S}_\mathrm{ns}$ span the image displacement subspace: this is why the mass-weighting here involves multiplying the displacement field $\mathbfcal{S}_\mathrm{ns}$ by $\mathbfcal{M}^{1/2}$ and not by $\mathbfcal{M}^{-1/2}$. This contrasts with the construction of polarization-space $\mathbfcal{A}$-type matrices, from the $\mathbfcal{L}$-type matrices, which introduce constraints of the form $\mathbfcal{L}^\intercal\mathbf{u}=0$.

\section{Bending modes}
\label{sec:bending}

Let us consider a vector $\vec r$ and the associated unit vector $\vec n=\vec r/r$. The orthogonal projection onto the plane perpendicular to $\vec n$ is given by: $P^\perp_{\vec n}=\mathbf{I}_3-\vec n\otimes\vec n$, where $\mathbf{I}_3$ denotes the $3\times3$ identity matrix. For any vector $\vec a$, the projection onto this plane, $\vec a^\perp=P^\perp_{\vec n}\cdot\vec a$, satisfies $\vec n\times(\vec n\times\vec a^\perp)=-\vec a^\perp$. As a consequence, $\vec n\times(\vec n\times\vec a)=-\vec a^\perp=-P^\perp_{\vec n}\cdot\vec a$, which constitutes an alternative expression for the projection $P^\perp_{\vec n}$.

Any variation of $\vec r$, induces a variation of the unit vector $\vec n=\vec r/r$ as $\d\vec n=(1/r)\,(1-\vec n\otimes\vec n)\cdot\d\vec r=(1/r)\,P^\perp_{\vec n}\cdot\d\vec r$. For any test vector $\vec a$, hence, $\vec a\cdot\d\vec n=(1/r)\,\vec a\cdot(P^\perp_{\vec n}\cdot\d\vec r)$. Using the symmetry of the projector and its alternative double-cross-product expression we obtain: $\vec a\cdot\d\vec n=(1/r)\,(P^\perp_{\vec n}\cdot\vec a)\cdot\d\vec r=-(1/r)\,(\vec n\times(\vec n\times\vec a))\cdot\d\vec r$.

Let us now consider the angle $\theta_{j_aij_b}=\arccos(\vec n_{ij_a}\cdot\vec n_{ij_b})\in[0,\pi]$ between any two vertices of a tetrahedron centered on silicon $i$. Its induced variation by any displacement field is calculated as follows:
\begin{equation}\label{eq:theta}
    \begin{split}
        \d\theta_{j_aij_b}&=-\frac{1}{\sin\theta_{j_aij_b}}\d\left(\vec n_{ij_a}\cdot\vec n_{ij_b}\right)\\
        &=-\frac{1}{\sin\theta_{j_aij_b}}\left(\vec n_{ij_b}\cdot\d\vec n_{ij_a}+\vec n_{ij_a}\cdot\d\vec n_{ij_b}\right)\\
        &=\frac{1}{\sin\theta_{j_aij_b}}\bigg(\left(\vec n_{ij_a}\times(\vec n_{ij_a}\times\vec n_{ij_b})\right)\cdot\frac{\d\vec r_{ij_a}}{r_{ij_a}}\\
        &\qquad\qquad+\left(\vec n_{ij_b}\times(\vec n_{ij_b}\times\vec n_{ij_a})\right)\cdot\frac{\d\vec r_{ij_b}}{r_{ij_b}}\bigg)\,,\\
    \end{split}
\end{equation}
where we have used the general relation $\vec a\cdot\d\vec n=-(1/r)\,(\vec n\times(\vec n\times\vec a))\cdot\d\vec r$ in the last step.

In the above expression, the elementary displacements $\d\vec r_{ij_a}$ and $\d\vec r_{ij_b}$ are contracted with  vectors of the form:
\begin{equation}
\begin{split}
\vec b_{j_aj_b}&\equiv-\frac{\vec n_{ij_a}\times(\vec n_{ij_a}\times\vec n_{ij_b})}{r_{ij_a}\,|\sin\theta_{j_aij_b}|}\\
&=\frac{1}{r_{ij_a}}\,\frac{(\vec r_{ij_a}\times\vec r_{ij_b})\times\vec r_{ij_a}}{\|(\vec r_{ij_a}\times\vec r_{ij_b})\times\vec r_{ij_a}\|}\\
\end{split}
\end{equation}
since $\|\vec r_{ij_a}\times(\vec r_{ij_a}\times\vec r_{ij_b})\|=r_{ij_a}\,\|\vec r_{ij_a}\times\vec r_{ij_b}\|=r_{ij_a}^2\,r_{ij_b}\,\sin\theta_{j_aij_b}$ (and likewise when swapping $a$ and $b$). The vector $\vec b_{j_aj_b}$ is perpendicular to $\vec r_{ij_a}$ and lies in the plane spanned by $(\vec r_{ij_a},\vec r_{ij_b})$. When anchored at $j_a$, it points towards the vertex $j_b$, i.e., represents an elementary displacement of $j_a$ that reduces $\theta_{j_aij_b}$. This choice of orientation (sign) follows a convention used in previous works~\cite{taraskin97,shcheblanov16}. In those studies, however, the vectors $\vec b_{j_aj_b}$ are taken to be normalized, which leads to elementary bending-mode expressions that implicitly assume a regular tetrahedral geometry. Here, to avoid introducing any unnecessary assumption, we instead define $\vec b_{j_aj_b}$ as above, with norm $\|\vec b_{j_aj_b}\|=1/r_{ij_a}$.

Having introduced this notation, the variation of the angle $\theta_{j_aij_b}$ obtained in Eq.~\eqref{eq:theta} can be recast as
\begin{equation}
    \d\theta_{j_aij_b}=-\vec b_{j_aj_b}\cdot\d\vec r_{ij_a}-\vec b_{j_bj_a}\cdot\d\vec r_{ij_b}\,.
\end{equation}
Let us now introduce the $3N$-dimensional coordinate vector $\mathbf{L}_{j_aj_b}^{(i)}$ whose only non-zero components are: $\vec L_{j_aj_b\ j_a}^{(i)}=\vec b_{j_aj_b}$ on oxygen $j_a$, $\vec L_{j_aj_b\ j_b}^{(i)}=\vec b_{j_bj_a}$ on oxygen $j_b$, and $\vec L_{j_aj_b\ i}^{(i)}=-\vec b_{j_aj_b}-\vec b_{j_bj_a}$ on the central silicon atom $i$. Denoting by $\d\mathbf{r}$ the generalized $3N$-dimensional elementary displacement vector, the angle variation can then be written in a compact form as
\begin{equation}\label{eq:dtheta}
\d\theta_{j_aij_b}=-\mathbf{L}_{j_aj_b}^{(i)}\cdot\d\mathbf{r}\qquad.
\end{equation}
The condition that the angle remains fixed reads $\mathbf{L}_{j_aj_b}^{(i)}\cdot\d\mathbf{r}=0$. Thus, $\mathbf{L}_{j_aj_b}^{(i)}=-\nabla_{\mathbf{r}}\theta_{j_aij_b}$ (minus the gradient of angle $\theta_{j_aij_b}$ with respect to the $3N$-component displacement field) generates an elementary contraction of $\theta_{j_aij_b}$.

There are six vertex pairs in a tetrahedron, and hence six associated angles. However, a simple geometric construction shows that once five of these angles are fixed, the sixth one is uniquely determined. A tetrahedron therefore possesses only five independent angular degrees of freedom, which parametrize its bending distortions.

Introducing the six-component vector $\boldsymbol{\theta}^{(i)}=\{\theta_{j_1ij_2},\theta_{j_1,i,j_3},\theta_{j_1ij_4},\theta_{j_2ij_3},\theta_{j_2ij_4},\theta_{j_3ij_4}\}$, a variation along the direction $\boldsymbol{\theta}_\mathrm{uni}^{(i)}=\{1,1,1,1,1,1\}$, corresponds to a uniform changes of all angles. In the particular case of a regular tetrahedron, since all six angles are equal and fixed to $\arccos(-1/3)$, any such variation is forbidden. Accordingly, bending coordinates are naturally defined by considering components of $\boldsymbol{\theta}^{(i)}$ in the five-dimensional hyperplane orthogonal to $\boldsymbol{\theta}_\mathrm{uni}^{(i)}$, which consists of linear combinations of angles the coefficients of which add up to zero.

To identify such combinations, we group the vertices into three pairs of complementary pairs, corresponding to opposite edges: $((j_1,j_2),(j_3,j_4))$, $((j_1,j_3),(j_2,j_4))$, $((j_1,j_4),(j_2,j_3))$. Then, we introduce the three differences of opposite angles, $\Delta_1^{(i)}=-\theta_{j_1ij_2}+\theta_{j_3ij_4}$, $\Delta_2^{(i)}=-\theta_{j_1ij_3}+\theta_{j_2ij_4}$, and $\Delta_3^{(i)}=-\theta_{j_1ij_4}+\theta_{j_2ij_3}$, and the corresponding sums, $S_1^{(i)}=\theta_{j_1ij_2}+\theta_{j_3ij_4}$, $S_2^{(i)}=\theta_{j_1ij_3}+\theta_{j_2ij_4}$, and $S_3^{(i)}=\theta_{j_1ij_4}+\theta_{j_2ij_3}$. These six quantities provide an alternative parametrization of the six bond angles. The differences $\Delta_1^{(i)}$, $\Delta_2^{(i)}$, and $\Delta_3^{(i)}$ describe coordinates along directions that are orthogonal to $\boldsymbol{\theta}_\mathrm{uni}^{(i)}$, but the sums do not. To construct components in the plane orthogonal to $\boldsymbol{\theta}_\mathrm{uni}^{(i)}$ from the sums, one must therefore combine them linearly in such a way that the coefficients add up to zero. This leaves two independent possibilities, which we choose as: $E_1^{(i)}=2S_1^{(i)}-S_2^{(i)}-S_3^{(i)}$ and $E_2^{(i)}=S_2^{(i)}-S_3^{(i)}$. Accordingly, the five independent bending degrees of freedom may be recast as the three differences of opposite angles $\Delta_1^{(i)}$, $\Delta_2^{(i)}$, and $\Delta_3^{(i)}$, together with the two coordinates $E_1^{(i)}$ and $E_2^{(i)}$, which combine sums of opposite angles while removing the uniform component.

The antisymmetric bending modes belonging to the representation $F_2^\mathrm{b}$ are associated with elementary variations of the three differences between opposite angles, $\Delta_1^{(i)}$, $\Delta_2^{(i)}$, and $\Delta_3^{(i)}$, respectively. Their generating vectors are immediately obtained by applying Eq.~\eqref{eq:dtheta} to the corresponding combinations of angles, yielding the expressions provided in Eq.~\eqref{eq:F2b}. The symmetric bending modes belonging to representation $E$ are associated with the variations of the two combinations $E_1^{(i)}$ and $E_2^{(i)}$; the corresponding generating vectors are similarly calculated and are provided in Eq.~\eqref{eq:E}.

For each irreducible subspace $G=F_2^\mathrm{b}$ or $E$, and for each silicon atom $i$, the local displacement field $\mathbf{L}_{G,m}^{(i)}$ generating the local bending mode $m$ has five non-zero atomic components: $\vec L_{G,m\,j_a}^{(i)}$ for each of the bonded oxygen atoms $j_a$ (with $a=1,\ldots,4$), and $\vec L_{G,m\,i}^{(i)}=-\sum_a\vec L_{G,m\,j_a}^{(i)}$ for the silicon atom. Each generating vector is thus fully specified by its four oxygen component vectors, and for simplicity we write $\mathbf{L}_{G,m}^{(i)}=\{\vec L_{G,m\,j_a}^{(i)},a=1,\ldots,4\}$.

Using this notation, the three generating vectors associated with $F_2^\mathrm{b}$ bending modes explicitly read
\begin{equation}
\begin{split}
  \mathbf{L}_{F_2^\mathrm{b},1}^{(i)}&=\{\vec b_{j_1j_2},\vec b_{j_2j_1},-\vec b_{j_3j_4},-\vec b_{j_4j_3}\} \,,\\
  \mathbf{L}_{F_2^\mathrm{b},2}^{(i)}&=\{\vec b_{j_1j_3},-\vec b_{j_2j_4},\vec b_{j_3j_1},-\vec b_{j_4j_2}\}\,,\\
  \mathbf{L}_{F_2^\mathrm{b},3}^{(i)}&=\{\vec b_{j_1j_4},-\vec b_{j_2j_3},-\vec b_{j_3j_2},\vec b_{j_4j_1}\}\,,
\end{split}
\end{equation}
while the two vectors generating $E$ modes are:
\begin{equation}
  \begin{split}
    \!\!\!\! \mathbf{L}_{E,1}^{(i)}\!=\!\{
    &2\vec b_{j_1j_2}-\vec b_{j_1j_3}-\vec b_{j_1j_4},
      2\vec b_{j_2j_1}-\vec b_{j_2j_3}-\vec b_{j_2j_4},\\
    &2\vec b_{j_3j_4}-\vec b_{j_3j_1}-\vec b_{j_3j_2},
      2\vec b_{j_4j_3}-\vec b_{j_4j_1}-\vec b_{j_4j_2}\}\,,\\
    \!\!\!\! \mathbf{L}_{E,2}^{(i)}\!=\!\{
    &\vec b_{j_1j_3}-\vec b_{j_1j_4},
      \vec b_{j_2j_4}-\vec b_{j_2j_3},\\
    &\vec b_{j_3j_1}-\vec b_{j_3j_2},
      \vec b_{j_4j_2}-\vec b_{j_4j_1}\} \, .
  \end{split}
\end{equation}

\bibliographystyle{apsrev4-2}
\bibliography{biblio}
\end{document}